\documentclass[prd,superscriptaddress,tightenlines,nofootinbib,
  eqsecnum]{revtex4}
\usepackage{amssymb,amsmath,amsfonts,graphicx,mathrsfs,color,bm}
\usepackage{empheq}
\usepackage{tikz}
\usepackage{tikz-cd}

\newcommand{\Comment}[1]{{}}
\definecolor{MyDarkBlue}{rgb}{0.15,0.15,0.45}
\usepackage[linktocpage=true]{hyperref}
\hypersetup{
colorlinks=true,
citecolor=MyDarkBlue,
linkcolor=MyDarkBlue,
urlcolor=MyDarkBlue,
pdfauthor={},
pdftitle={},
pdfsubject={}
}

\allowdisplaybreaks
\usepackage{ulem}
\newcommand{\be}{\begin{equation}}
\newcommand{\ee}{\end{equation}}
\newcommand{\sbe}{\begin{subequations}}
\newcommand{\see}{\end{subequations}}

\newcommand{\ba}{\begin{eqnarray}}
\newcommand{\ea}{\end{eqnarray}}
\newcommand{\p}{\partial}
\newcommand{\nn}{\nonumber}
\newcommand{\ud}{\mathrm{d}}
 
\newcommand{\ue}{\mathrm{e}}
\newcommand\calO{{\mathcal{O}}}

\begin{document}

\title{Dynamics of compact binary systems in scalar-tensor theories: I. Equations of motion to the third post-Newtonian order}

\author{Laura Bernard}\email{laura.bernard@tecnico.ulisboa.pt}
\affiliation{CENTRA, Departamento de F\'{\i}sica, Instituto Superior T{\'e}cnico -- IST, Universidade de Lisboa -- UL, Avenida Rovisco Pais 1, 1049 Lisboa, Portugal}

\date{\today}

\begin{abstract}

Scalar-tensor theories are one of the most natural and well-constrained alternative theories of gravity, while still allowing for significant deviations from general relativity. We present the equations of motion of nonspinning compact binary systems at the third post-Newtonian (PN) order in massless scalar-tensor theories. We adapt the Fokker action of point particles in harmonic coordinates in general relativity to the specificities of scalar-tensor theories. We use dimensional regularisation to treat both the infrared and ultraviolet divergences, and we consistently include the tail effects that contribute by a non-local term to the dynamics. This work is crucial in order to later compute the scalar gravitational waveform and the energy flux at 2PN order.

\end{abstract}

\maketitle

\section{Introduction}\label{sec:intro}

The observations by the LIGO-Virgo collaboration of gravitational waves emitted by coalescing compact binary systems have opened a new era in gravitational wave astronomy~\cite{Abbott:2016blz,Abbott:2016nmj,Abbott:2017vtc,Abbott:2017oio,TheLIGOScientific:2017qsa}. In the upcoming year, we expect to see many of these events, both in the advanced earth-based interferometric detectors, and in the space-based antenna LISA. The gravitational wave observations will allow us to not only measure the astrophysical properties of these systems, but also to challenge general relativity (GR) in the strong-field and highly dynamical regime of gravity.

The detection and parameter estimation of gravitational wave events require a bank of highly accurate templates for the gravitational waveforms. For the inspiral part of the coalescence of compact binary systems, the post-Newtonian formalism is well-suited to describe the evolution of the system~\cite{Blanchet:2013haa}: it consists of an expansion in the small parameter ${\varepsilon\equiv v/c\sim \left(Gm/rc^{2}\right)^{1/2}}$. The current state of the art in GR concerning the dynamics is the 4PN order\footnote{As usual, we refer to post-Newtonian order as $n\mathrm{PN}\equiv\mathcal{O}\left(v^{2}/c^{2}\right)^{n}$.}~\cite{Foffa:2012rn,Jaranowski:2012eb,Jaranowski:2013lca,Damour:2014jta,Bernard:2015njp,Bernard:2016wrg,Bernard:2017bvn,Marchand:2017pir}. The energy flux is known up to $3.5$PN order beyond the quadrupole formula~\cite{Blanchet:1995ez,Blanchet:1997jj,Blanchet:2001ax,Blanchet:2004ek}, with the 4.5PN coefficient also being known~\cite{Marchand:2016vox}; while the dominant modes of the gravitational waveform are known up to 3.5PN order~\cite{Blanchet:2008je,Faye:2012we,Faye:2014fra}. The complete waveform is obtained by connecting the PN result with numerical relativity waveforms. At present, this is done using either a direct matching (IMR models)~\cite{Ajith:2009bn} or some resummation techniques (EOB waveforms)~\cite{Buonanno:1998gg}.

In order to test general relativity, one also has to model waveforms in alternative theories of gravity. Existing tests are performed using either theory-independent or theory-dependent methods. In this paper, we focus on a particular class of theories, namely massless scalar-tensor (ST) theories, which are among the most popular and well-studied theories. They date back to more than sixty years ago, when they were introduced by Jordan, Fierz, Brans and Dicke. See~\cite{Brans:2008zz,Fujii:2005} for historical reviews of these theories and~\cite{Esposito-Farese:2011cha} for current constraints on the parameters. One of the motivations for studying these theories is to explain the accelerated expansion of the universe, as $f(R)$-theories, in which the action is expressed as a function of the Ricci scalar, can be expressed as a scalar-tensor theory~\cite{DeFelice:2010aj}.

Previous works in order to obtain the waveform at 2PN order have been performed during the last five years. The equations of motion are known at $2.5$PN order~\cite{Mirshekari:2013vb}, while the tensor gravitational waveform is known at 2PN order~\cite{Lang:2013fna}. However, the scalar waveform is only known at $1.5$PN and the energy flux at $1$PN order, as they respectively start at $-0.5$PN and $-1$PN order with respect to the leading GR order~\cite{Lang:2014osa}. All these ST results were obtained using the direct integration of relaxed Einstein equations (DIRE) method developped by Will, Wiseman and Pati~\cite{Wiseman:1992dv,Will:1996zj,Pati:2000vt}. The ``Effective One-Body formalism'' (EOB) has also been developped for ST theories, focusing on the derivation of a ST-EOB Hamiltonian~\cite{Julie:2017pkb,Julie:2017ucp}. Numerical works have shown that compact binaries in scalar-tensor theories can undergo a dynamical scalarisation phenomenon~\cite{Barausse:2012da,Palenzuela:2013hsa}, similar to the spontaneous scalarisation effect for individual stars~\cite{Damour:1993hw,Damour:1996ke}. This phenomenon happens during the late-inspiral phase, where the post-Newtonian approximation is expected to break done. Recently, an analytical method has been proposed to capture dynamical scalarisation, using resummation techniques~\cite{Sennett:2016rwa}.

In order to compute the scalar waveform and energy flux at 2PN order, the equations of motion at 3PN order are required. In the present paper, we pursue this aim by constructing a Fokker action of point particles in harmonic coordinates. This method has recently been developped to successfully derive the 4PN equations of motion in GR~\cite{Bernard:2015njp}. Here, we adapt this approach to the specificities of scalar-tensor theories. We use dimensional regularisation to treat both the infrared and ultraviolet divergences. We show that some tail effects appear at 3PN in ST theories, associated to the scalar dipole moment, while these effects start contributing only at 4PN in GR~\cite{Blanchet:1993ng}. We then obain a complete ambiguity-free result, as expected from the recent computation at 4PN in GR~\cite{Bernard:2017bvn,Marchand:2017pir}. In the companion paper~\cite{Bernard:2018}, we will study the conserved integrals of motion and the reduction to the center-of-mass frame.

In the following, we present in section~\ref{sec:STtheory} our massless scalar-tensor theory, and derive the equations of motion. In section~\ref{sec:MPNinST}, we adapt the multipolar post-Newtonian formalism to ST theories. In particular, we consistently incorporate the tail effects that contribute to the conservative 3PN dynamics. In section~\ref{sec:Fokker}, we implement the post-Newtonian solution into the Fokker action, and explain the dimensional regularisation method. Finally in section~\ref{sec:results}, we show the full 3PN equations of motion in harmonic coordinates for ST theories and conclude with some comments on our result.

\paragraph*{\textbf{Notations:}}

We use boldface letter to represent three-dimensional Euclidean vectors. We denote by $\mathbf{y}_{A}(t)$ the two ordinary coordinate trajectories in a harmonic coordinate system $\left\{t,\mathbf{x}\right\}$, by $\mathbf{v}_{A}(t)=\ud\mathbf{y}_{A}/\ud t$ the two ordinary velocities and by $\mathbf{a}_{A}(t)=\ud\mathbf{v}_{A}/\ud t$ the two ordinary accelerations. The ordinary separation vector reads $\mathbf{n}_{12}=\left(\mathbf{y}_{1}-\mathbf{y}_{2}\right)/r_{12}$, where $r_{12}=\left\vert\mathbf{y}_{1}-\mathbf{y}_{2}\right\vert$. Ordinary scalar products are denoted, \textit{e.g.} $\left(n_{12}v_{1}\right)=\mathbf{n}_{12}\cdot\mathbf{v}_{1}$, while the two masses are indicated by $m_{1}$ and $m_{2}$. We note $\hat{n}_{L}$ the
symmetric trace-free (STF) product of $\ell$ spatial vectors $n_{i}$, with $L=i_{1}\cdots i_{l}$ a multi-index made of $\ell$ spatial indices.

\section{Massless scalar-tensor theories}\label{sec:STtheory}

\subsection{The field equations in ST theories}\label{subsec:rEFE}

We consider a generic class of massless scalar-tensor theories in which a single massless scalar field $\phi$ minimally couples to the metric $g_{\mu\nu}$. It is described by the action
\be\label{STactionJF}
S_{\mathrm{st}} = \frac{c^{3}}{16\pi G} \int\ud^{4}x\,\sqrt{-g}\left[\phi R - \frac{\omega(\phi)}{\phi}g^{\alpha\beta}\p_{\alpha}\phi\p_{\beta}\phi\right] +S_{\mathrm{m}}\left(\mathfrak{m},g_{\alpha\beta}\right)\,,
\ee
where $R$ and $g$ are respectively the Ricci scalar and the determinant of the metric, $\omega$ is a function of the scalar field and $\mathfrak{m}$ stands generically for the matter fields. The action for the matter $S_{\mathrm{m}}$ is a function only of the matter fields and the metric. The action~\eqref{STactionJF} is often called the ``metric'' or ``Jordan''-frame action, as the matter does not couple directly to the scalar field.

We note $\phi_{0}$ the value of the scalar field at spatial infinity and we assume that it is constant in time. We then define the rescaled scalar field $\varphi\equiv \frac{\phi}{\phi_{0}}$ and the conformally related metric,
\be
\tilde{g}_{\mu\nu}\equiv \varphi\,g_{\mu\nu}\,.
\ee
In terms of these new variables, the action~\eqref{STactionJF} can be rewritten as,
\be\label{STactionEF}
S_{\mathrm{st}} = \frac{c^{3}\phi_{0}}{16\pi G} \int\ud^{4}x\,\sqrt{-\tilde{g}}\left[ \tilde{R} + \frac{3}{\varphi}\tilde{g}^{\alpha\beta}\nabla_{\alpha}\p_{\beta}\varphi - \frac{9+2\omega(\phi)}{2\varphi^{2}}\tilde{g}^{\alpha\beta}\p_{\alpha}\varphi\p_{\beta}\varphi\right] +S_{\mathrm{m}}\left(\mathfrak{m},g_{\alpha\beta}\right)\,,
\ee
Note that the matter fields still couple to the physical metric $g_{\mu\nu}$. As the scalar field is now minimally coupled to the metric, the action~\eqref{STactionEF} is often called the ``Einstein''-frame action, and we will do our calculation in this frame.
Next, we perforn some integrations by part to rewrite the action~\eqref{STactionEF} into the Landau-Lifshitz form and we insert a harmonic gauge-fixing term $-\frac{1}{2}\tilde{g}_{\mu\nu}\tilde{\Gamma}^{\mu}\tilde{\Gamma}^{\nu}$. The new action is fully equivalent to the previous one and reads,
\be\label{STactionEFgf}
S_{\mathrm{ST}} = \frac{c^{3}\phi_{0}}{16\pi G} \int\ud^{4}x\,\sqrt{-\tilde{g}}\left[ \tilde{g}^{\mu\nu}\left(\tilde{\Gamma}^{\rho}_{\mu\lambda}\tilde{\Gamma}^{\lambda}_{\nu\rho}- \tilde{\Gamma}^{\rho}_{\mu\nu}\tilde{\Gamma}^{\lambda}_{\rho\lambda}\right) -\frac{1}{2}\tilde{g}_{\mu\nu}\tilde{\Gamma}^{\mu}\tilde{\Gamma}^{\nu} - \frac{3+2\omega(\phi)}{2\varphi^{2}}\tilde{g}^{\alpha\beta}\p_{\alpha}\varphi\p_{\beta}\varphi\right] +S_{\mathrm{m}}\left(\mathfrak{m},g_{\alpha\beta}\right)\,,
\ee
where $\tilde{\Gamma}^{\mu}\equiv \tilde{g}^{\rho\sigma}\tilde{\Gamma}^{\mu}_{\rho\sigma}$ and $\tilde{\Gamma}^{\mu}_{\rho\sigma}$ are the Christoffel symbols of the conformal metric. Defining the inverse gothic metric by
\be
\mathfrak{\tilde{g}}^{\mu\nu}\equiv\sqrt{-\tilde{g}}\tilde{g}^{\mu\nu}\,,
\ee
the action~\eqref{STactionEF} can further be rewritten as
\begin{align}\label{STactionEFgot}\nn
S_{\mathrm{ST}} = \frac{c^{3}\phi_{0}}{32\pi G} \int\ud^{4}x\,\biggl[& -\frac{1}{2}\left(\tilde{\mathfrak{g}}_{\mu\rho}\tilde{\mathfrak{g}}_{\nu\sigma} -\frac{1}{2}\tilde{\mathfrak{g}}_{\mu\nu}\tilde{\mathfrak{g}}_{\rho\sigma}\right) \tilde{\mathfrak{g}}^{\lambda\tau}\partial_{\lambda}\tilde{\mathfrak{g}}^{\mu\nu}\partial_{\tau} \tilde{\mathfrak{g}}^{\rho\sigma} \\
& +\tilde{\mathfrak{g}}_{\mu\nu}\Bigl(\partial_{\rho}\tilde{\mathfrak{g}}^{\mu\sigma}  \partial_{\sigma}\tilde{\mathfrak{g}}^{\nu\rho} -\partial_{\rho}\tilde{\mathfrak{g}}^{\mu\rho}\partial_{\sigma}\tilde{\mathfrak{g}}^{\nu\sigma}\Bigr) - \frac{3+2\omega}{\varphi^{2}}\tilde{\mathfrak{g}}^{\alpha\beta}\p_{\alpha}\varphi\p_{\beta}\varphi\biggr] +S_{\mathrm{m}}\left(\mathfrak{m},g_{\alpha\beta}\right)\,.
\end{align}
Next, we expand the gothic metric around Minkowski space-time and define the metric and scalar perturbation variables $h^{\mu\nu}$ and $\psi$ by
\be
h^{\mu\nu}\equiv \mathfrak{\tilde{g}}^{\mu\nu}-\eta^{\mu\nu}\,, \qquad\text{and}\qquad \psi\equiv\varphi-1\,.
\ee
The field equations derived from the gauge-fixed action~\eqref{STactionEFgot} read,
\begin{subequations}\label{rEFE}
\begin{align}
& \square_{\eta}\,h^{\mu\nu} = \frac{16\pi G}{c^{4}}\tau^{\mu\nu}\,,\\
& \square_{\eta}\,\psi = -\frac{8\pi G}{c^{4}}\tau_{s}\,,
\end{align}
\end{subequations}
with
\begin{subequations}
\begin{align}\label{taumunu}
& \tau^{\mu\nu} = \frac{\varphi}{\phi_{0}}\left[(-g)T^{\mu\nu}\right] +\frac{c^{4}}{16\pi G}\Sigma^{\mu\nu}+\frac{c^{4}}{16\pi G}\Lambda_{\mathrm{S}}^{\mu\nu}\,,\\
\label{taus} & \tau_{s} = -\frac{\varphi}{\phi_{0}(3+2\omega)}\sqrt{-g}\left(T-2\varphi\frac{\p T}{\p \varphi}\right) -\frac{c^{4}}{8\pi G}\left[-h^{\alpha\beta}\p_{\alpha}\p_{\beta}\psi-\p_{\alpha}\psi\p_{\beta}h^{\alpha\beta} +\left(\frac{1}{\varphi}-\frac{\phi_{0}\omega'}{3+2\omega}\right)\tilde{\mathfrak{g}}^{\alpha\beta}\p_{\alpha}\varphi\p_{\beta}\varphi\right] \,,
\end{align}
\end{subequations}
where $T^{\mu\nu}=\frac{2}{\sqrt{-g}}\frac{\delta S_{\mathrm{m}}}{\delta g_{\mu\nu}}$ is the matter stress-energy tensor and $T\equiv g_{\mu\nu}T^{\mu\nu}$. The scalar source term $\Lambda_{\mathrm{S}}^{\mu\nu}$ is given by
\be\label{Lambdas}
\Lambda_{\mathrm{S}}^{\mu\nu} = \frac{3+2\omega}{\varphi^2}\left(\tilde{\mathfrak{g}}^{\mu\alpha}\tilde{\mathfrak{g}}^{\nu\beta} -\frac{1}{2}\tilde{\mathfrak{g}}^{\mu\nu}\tilde{\mathfrak{g}}^{\alpha\beta}\right)\p_{\alpha}\varphi\p_{\beta}\varphi\,.
\ee
The gravitational source term $\Sigma^{\mu\nu}=\Lambda^{\mu\nu}_{\mathrm{LL}}+\Lambda_{\mathrm{H}}^{\mu\nu}+\Lambda^{\mu\nu}_{\mathrm{gf}}$, where $\Lambda^{\mu\nu}_{\mathrm{LL}}$ is the Landau-Lifshitz pseudo-energy tensor~\cite{Landau1971}, is at least quadratic in the field $h$ and its derivatives, with components given by
\begin{subequations}
\begin{align}\label{LambdaLL}
& \Lambda_{\mathrm{LL}}^{\alpha\beta} =\ \frac{1}{2}\tilde{\mathfrak{g}}^{\alpha\beta}\tilde{\mathfrak{g}}_{\mu\nu}\partial_{\lambda}h^{\mu\gamma}\partial_{\gamma}h^{\nu\lambda}-\tilde{\mathfrak{g}}^{\alpha\mu}\tilde{\mathfrak{g}}_{\nu\gamma}\partial_{\lambda}h^{\beta\gamma}\partial_{\mu}h^{\nu\lambda} -\tilde{\mathfrak{g}}^{\beta\mu}\tilde{\mathfrak{g}}_{\nu\gamma}\partial_{\lambda}h^{\alpha\gamma}\partial_{\mu}h^{\nu\lambda} \nn&\\
&\qquad\quad +\tilde{\mathfrak{g}}_{\mu\nu}\tilde{\mathfrak{g}}^{\lambda\gamma}\partial_{\lambda}h^{\alpha\mu}\partial_{\gamma}h^{\beta\nu} +\frac{1}{8}\left(2\tilde{\mathfrak{g}}^{\alpha\mu}\tilde{\mathfrak{g}}^{\beta\nu}-\tilde{\mathfrak{g}}^{\alpha\beta}\tilde{\mathfrak{g}}^{\mu\nu}\right)\left(2\tilde{\mathfrak{g}}_{\lambda\gamma}\tilde{\mathfrak{g}}_{\tau\pi}-\tilde{\mathfrak{g}}_{\gamma\tau}\tilde{\mathfrak{g}}_{\lambda\pi}\right)\partial_{\mu}h^{\lambda\pi}\partial_{\nu}h^{\gamma\tau} \,, \\
\label{LambdaH}
&\Lambda_{\mathrm{H}}^{\alpha\beta} = -h^{\mu\nu}\partial_{\mu}\partial_{\nu}h^{\alpha\beta} +\partial_{\mu}h^{\alpha\nu}\partial_{\nu}h^{\beta\mu} \,,\\
\label{Lambdagf}
& \Lambda_{\mathrm{gf}}^{\alpha\beta} = - \partial_{\lambda}h^{\lambda\alpha} \partial_{\sigma}h^{\sigma\beta} - \partial_{\lambda}h^{\lambda\rho}\partial_{\rho}h^{\alpha\beta} -\frac{1}{2}\tilde{\mathfrak{g}}^{\alpha\beta}\tilde{\mathfrak{g}}_{\rho\sigma} \partial_{\lambda}h^{\lambda\rho} \partial_{\gamma}h^{\gamma\sigma} +2\tilde{\mathfrak{g}}_{\rho\sigma}\tilde{\mathfrak{g}}^{\lambda(\alpha} \partial_{\lambda}h^{\beta)\rho}\partial_{\gamma}h^{\gamma\sigma} \,.
\end{align}
\end{subequations}
Note that the gauge-fixing term~\eqref{Lambdagf} contains the harmonicities $\p_{\nu}h^{\mu\nu}$ which are not zero in general. However, this term will ensure that, on-shell, our results are in harmonic coordinates.

\subsection{The action for matter}\label{subsec:matter}

We now make precise the action describing the matter. As we are dealing with compact, self-gravitating objects in scalar-tensor theories, we have to take into account the internal gravity of each body. To do so, we follow the approach pioneered by Eardley~\cite{Eardley1975} and consider that the total mass of each body may depend on the value of the scalar field at its location. The skeletonized matter action is then given by the classical action for point particles, but with a mass $m_{A}(\phi)$, namely
\be\label{matteract}
S_{\mathrm{m}} = -\sum_{A}\int\ud t\,m_{A}(\phi)c^{2}\sqrt{-\left(g_{\alpha\beta}\right)_{A}\frac{v_{A}^{\alpha}v_{A}^{\beta}}{c^2}}\,,
\ee
where $v_{A}^{\mu}\equiv\frac{\ud y_{A}^{\mu}}{\ud t}=\left(c,\mathbf{v}_{A}\right)$ is the coordinate velocity of particle $A$, $y_{A}^{\mu}=\left(ct,\mathbf{y}_{A}\right)$ its trajectory and $\left(g_{\alpha\beta}\right)_{A}$ is the physical metric evaluated at the position of particle $A$ using the dimensional regularisation scheme. We recall that the physical metric is related to the conformal one through $g_{\alpha\beta}=\frac{\tilde{g}_{\alpha\beta}}{\phi}$. Note that the scalar-field dependence of the mass is responsible for the term $\frac{\p T}{\p \varphi}$ in Eq.~\eqref{taus}. In the absence of such a dependence, \textit{e.g.} in GR, the matter stress-energy tensor should depend only on the matter variables and the metric.

We then define the sensitivity of each body with respect to the scalar field as
\be
s_{A} \equiv \left.\frac{\ud\ln m_{A}(\phi)}{\ud\ln\phi}\right\vert_{\phi={\phi_{0}}} \,.
\ee
In the calculation at 3PN, we will also need the higher order sensitivities, defined in Sec.~\ref{sec:results}. The sensitivities of neutron stars are around $s_{\mathrm{NS}}\sim 0.2$, depending on the mass and the equation of state. Due to dynamical scalarisation, neutron star sensitivities can dramatically grow during the late-inspiral. As we are working in the post-Newtonian formalism and we assume that the sensitivities are constant, we will not describe this effect in our work.  
Hawking's theorem states that stationary black holes have no hair in Brans-Dicke theory~\cite{Hawking:1972qk} and this result has been extended to generalised scalar-tensor theories~\cite{Sotiriou:2011dz}. Thus, for stationary black holes, the sensitivity is exactly $s_{\mathrm{BH}}=\frac{1}{2}$. Another way to see it is to define the scalar charges~\cite{Damour:1992we,Damour:1993hw},
\be
\alpha_{A} \equiv \frac{1-2s_{A}}{\sqrt{3+2\omega_{0}}}\,,
\ee
where $\omega_{0}\equiv\omega(\phi_{0})$. We see that $s_{\mathrm{BH}}=\frac{1}{2}$ implies $\alpha_{\mathrm{BH}}=0$, i.e. that stationary black holes have no hair. However, in the case of non-stationnary black-holes, i.e. for a time-varying scalar background, it has been shown that a scalar hair can arise~\cite{Jacobson:1999vr,Horbatsch:2011ye}. A similar result has been obtain in the presence of a constant scalar gradient in the background~\cite{Berti:2013gfa}.

\subsection{The Fokker action}\label{subsec:Fokker}

The Fokker action is then computed by replacing into the original action the gravitational and scalar degrees of freedom by their solution, obtained by resolving the field equations~\eqref{rEFE},
\be\label{Fokker}
S_{\mathrm{Fokker}}\left[\mathbf{y}_{A}(t),\mathbf{v}_{A}(t),\cdots\right] \equiv S_{\mathrm{ST}}\left[\tilde{g}^{\mathrm{(sol)}}_{\mu\nu}\left(\mathbf{y}_{B}(t),\mathbf{v}_{B}(t),\cdots\right),\varphi^{\mathrm{(sol)}}\left(\mathbf{y}_{B}(t),\mathbf{v}_{B}(t),\cdots\right),\mathbf{v}_{A}(t)\right]\,.
\ee
This procedure only applies to the conservative dynamics\footnote{An effective field theory method to compute the dissipative effects in the dynamics from a Lagrangian, consisting in doubling the matter variables, has been developped for GR~\cite{Galley:2015kus}}. In general relativity, starting at 2PN order, the Lagrangian depends linearly in the accelerations~\cite{Damour1981}, and as expected, we recover this feature in ST theories~\cite{Mirshekari:2013vb}. At 3PN order, we first obtain a Lagrangian that also contains terms quadratic or of higher order in the accelerations and derivatives of the accelerations. By implementing the \textit{double-zero} method~\cite{Damour:1985mt} and adding total time-derivatives, that do not contribute to the dynamics, we can reduce our original result to a Lagrangian linear in the accelerations. The equations of motion for the particles are then obtained by writing the genelarised Euler-Lagrange equations,
\be\label{GenLagEqs}
\frac{\delta S_{\mathrm{Fokker}}}{\delta\mathbf{y}_{A}} \equiv \frac{\p L_{\mathrm{F}}}{\p\mathbf{y_{A}}}-\frac{\ud}{\ud t}\left(\frac{\p L_{\mathrm{F}}}{\p\mathbf{v_{A}}}\right)+\frac{\ud^{2}}{\ud t^{2}}\left(\frac{\p L_{\mathrm{F}}}{\p\mathbf{a_{A}}}\right) +\cdots\,,
\ee
where $L_{\mathrm{F}}$ is the Lagrangian corresponding to the action, $S_{\mathrm{Fokker}}=\int\ud t\,L_{\mathrm{F}}$. Only once we have constructed the equations of motion using Eq.~\eqref{GenLagEqs}, do we order reduce them by replacing the accelerations by their lower order value.

\section{The Multipolar post-Newtonian formalism in scalar-tensor theories}\label{sec:MPNinST}

\subsection{The separation between the near and wave zones}

We generically denote $\left(\overline{h},\overline{\psi}\right)$ the PN solution of the field equations in the near-zone of the compact source, i.e. in a region of small extent compared to the gravitational wavelength. It is obtained by a PN iteration of the field equations~\eqref{rEFE}. In the exterior region of the source, including the wave zone, the multipolar solution is obtained by a post-Minkowskian iteration of the field equations in vacuum and is denoted $\mathcal{M}\left(h,\psi\right)$. As we are dealing with a post-Newtonian source, i.e. a compact weakly-stressed and slowly moving source, there exists a buffer region where the two expansions are valid. The complete solution is then obtained by a careful matching of the two solutions in the exterior part of the near zone, using the method of matched asymptotic expansions~\cite{Blanchet:2013haa}. In particular, we impose the matching equation,
\be\label{MatchingEq}
\overline{\mathcal{M}\left(h,\psi\right)} = \mathcal{M}\left(\overline{h},\overline{\psi}\right) \,,
\ee
i.e. that the multipolar expansion of the PN solution is equal to the PN expansion of the multipolar solution. We emphasize that Eq.~\eqref{MatchingEq} is valid everywhere and not only in the buffer zone. The carefull implementation of Eq.~\eqref{MatchingEq} is crucial when calculating the tail contribution to the 3PN equations of motion.

The gravitational part of the action $S_{\mathrm{g}}=\int\ud t\,L_{\mathrm{g}}$ can be decomposed according to
\be\label{lemma1}
L_{\mathrm{g}} = \int\ud^{d}x\,\overline{\mathcal{L}_{\mathrm{g}}} + \int\ud^{d}x\,\mathcal{M}\left(\mathcal{L}_{\mathrm{g}}\right) \,,
\ee
where $\mathcal{L}_{\mathrm{g}}$ is the Lagrangian density. We use dimensional regularisation to treat the infrared divergences of the post-Newtonian solution at infinity and the ultraviolet divergences of the multipolar solution at zero. The proof of this equation can be found in Appendix~\ref{sec:Lemma1}\footnote{The proof is similar to the one that one can find in section~(II.B) of~\cite{Bernard:2015njp}. The only difference lies in the fact that we are now dealing with dimensional regularisation while in~\cite{Bernard:2015njp}, the proof was done using a Hadamard-type regularisation.}. It uses the formal structure of the multipolar expansion ${\mathcal{M}\left(\mathcal{L}_{\mathrm{g}}\right)\sim\sum\hat{n}_{L}r^{a}(\ln r)^{b}F(t)}$ and the fact that the integral over space of such generic terms is always zero by analytic continuation in $\varepsilon\equiv d-3$. Next, we investigate the second term in~\eqref{lemma1}. In~\cite{Bernard:2015njp}, it was shown that this integral is zero for instantaneous terms, namely
\be\label{lemma2}
\int\ud^{d}x\,\left.\mathcal{M}\left(\mathcal{L}_{\mathrm{g}}\right)\right\vert_{\mathrm{inst}} = 0 \,.
\ee
Thus, the only contributions come from hereditary terms, that have the formal structure
\be\label{heredterm}
\left.\mathcal{M}\left(\mathcal{L}_{\mathrm{g}}\right)\right\vert_{\mathrm{hered}} = \sum\frac{\hat{n}_{L}}{r^{k}}(\ln r)^{q}H(u)\int_{-\infty}^{u}\ud v\,\mathcal{Q}\left(1+\frac{u-v}{r}\right)K(v) \,,
\ee
where $u=t-r/c$ is the retarded time, and $H$ and $K$ are functions of the source multipole moments $I_{L}$ and $J_{L}$. In ST theories, the multipole expansion of the Lagrangian density has the following formal structure after some integrations by part,
\be\label{Lagdens}
\mathcal{M}\left(\mathcal{L}_{\mathrm{g}}\right) \sim \mathcal{M}(h)\square\mathcal{M}(h) +\mathcal{M}(\psi)\square\mathcal{M}(\psi) +\mathcal{M}(h,\psi)\p\mathcal{M}(h,\psi)\p\mathcal{M}(h,\psi) +\cdots\,.
\ee
As $\mathcal{M}(h)$ and $\mathcal{M}(\psi)$ are solutions of the vacuum field equations, their source is at least quadratic in the fields, that is
\be\label{source}
\square\mathcal{M}(h) \sim \p\mathcal{M}(h,\psi)\p\mathcal{M}(h,\psi)\,, \qquad\text{and}\qquad \square\mathcal{M}(\psi) \sim \p\mathcal{M}(h,\psi)\p\mathcal{M}(h,\psi) \,.
\ee
Inserting Eq.~\eqref{source} into  Eq.~\eqref{Lagdens} we see that $\mathcal{M}\left(\mathcal{L}_{\mathrm{g}}\right)$ is at least cubic in the gravitational fields, and will be at least of order $\mathcal{O}\left(G^{3}\right)$. As we know that $\mathcal{M}\left(\mathcal{L}_{\mathrm{g}}\right)$ should contain at least one hereditary term, the dominant effect corresponds to an interaction of the type $M\times M\times I_{L}$, the so-called ``tails-of-tails''. In GR, when the scalar field is absent, these terms arise at least at $5.5$PN order corresponding to an interaction between two mass-monopoles and one mass-quadrupole~\cite{Blanchet:2013txa}. In ST theories, in addition to this effect we can also have an interaction between two mass-monopoles and one scalar mass-dipole, giving a first contribution at $4.5$PN order. We conclude that the second term in the r.h.s of Eq.~\eqref{lemma1} is at least of order $4.5$PN, and will not contribute to the dynamics at $3$PN order.

Thus, the gravitational part of the Lagrangian has to be computed only using the PN solution only, namely
\be\label{LagPN}
L_{\mathrm{g}} = \int\ud^{d}x\,\overline{\mathcal{L}_{\mathrm{g}}}\,.
\ee
The post-Newtonian solutions $\left(\overline{h},\overline{\psi}\right)$, obtained by solving the field equations~\eqref{rEFE}, read
\sbe\label{PNsol}
\begin{align}
& \overline{h}^{\mu\nu} = \frac{16\pi G}{c^{4}} \overline{\square_{\mathrm{ret}}^{-1}}\left[r^{\eta}\overline{\tau}^{\mu\nu}\right] + \mathcal{H}^{\mu\nu} \,,\\
& \overline{\psi} = -\frac{8\pi G}{c^{4}} \overline{\square_{\mathrm{ret}}^{-1}}\left[r^{\eta}\overline{\tau}_{\mathrm{s}}\right] + \Psi\,,
\end{align}
\see
where an overline denotes a PN expansion. The first terms in Eqs.~\eqref{PNsol} are particular retarded solutions of the PN-expanded field equations~\eqref{rEFE}. They read
\sbe\label{PNpartsol}
\begin{align}
& \overline{\square_{\mathrm{ret}}^{-1}}\left[r^{\eta}\overline{\tau}^{\mu\nu}\right] = -\frac{\tilde{k}}{4\pi}\int\ud^{d}\mathbf{x'}\vert\mathbf{x'}\vert^{\eta}\overline{\int_{1}^{+\infty}\ud z\,\gamma_{\tfrac{1-d}{2}}(z)\frac{\overline{\tau}^{\mu\nu}\left(\mathbf{x'},t-z\vert\mathbf{x}-\mathbf{x'}\vert/c\right)}{\vert\mathbf{x}-\mathbf{x'}\vert^{d-2}}} \,,\\
& \overline{\square_{\mathrm{ret}}^{-1}}\left[r^{\eta}\overline{\tau}_{\mathrm{s}}\right] = -\frac{\tilde{k}}{4\pi}\int\ud^{d}\mathbf{x'}\vert\mathbf{x'}\vert^{\eta}\overline{\int_{1}^{+\infty}\ud z\,\gamma_{\tfrac{1-d}{2}}(z)\frac{\overline{\tau}_{\mathrm{s}}\left(\mathbf{x'},t-z\vert\mathbf{x}-\mathbf{x'}\vert/c\right)}{\vert\mathbf{x}-\mathbf{x'}\vert^{d-2}}} \,,
\end{align}
\see
where $\tilde{k}=\frac{\Gamma\left(\frac{d}{2}-1\right)}{\pi^{\frac{d}{2}-1}}$, $\Gamma$ is the Eulerian function, and the function $\gamma_{\frac{1-d}{2}}(z)$ is defined by
\be
\gamma_{s}(z)=\frac{2\sqrt{\pi}}{\Gamma\left(s+1\right)\Gamma\left(-s-\tfrac{1}{2}\right)}\left(z^{2}-1\right)^{s}\,,
\ee
with the normalisation $\int_{1}^{+\infty}\gamma_{s}(z)=1$. The retarded Green's function of the scalar wave equation $G_{\mathrm{ret}}\left(\mathbf{x},t\right)$, solution of $\square G_{\mathrm{ret}}=\delta(t)\delta^{(d)}(\mathrm{x})$, is then given by
\be
G_{\mathrm{ret}}\left(\mathbf{x},t\right) = -\frac{\tilde{k}}{4\pi}\frac{\theta(t-r)}{r^{d-1}}\gamma_{\frac{1-d}{2}}\left(\frac{t}{r}\right)\,,
\ee
where $\theta(t-r)$ is the usual Heaviside step function.
In Eq.~\eqref{PNpartsol}, we have used the so-called ``$\varepsilon\eta$'' regularisation scheme, which is the equivalent for dimensional regularisation of the finite part procedure of Hadamard regularisation. It has recently been successfully used to compute the ambiguities at 4PN in GR~\cite{Bernard:2017bvn,Marchand:2017pir}. We have introduced a factor $r^{\eta}$ multiplying the PN source term, that acts as a regulator acting on top of dimensional regularisation. In practice, we shall first take the limit $\eta\rightarrow 0$ in generic $d$ dimensions and then take the limit $\varepsilon=d-3\rightarrow 0$. Although some poles in $1/\eta$ may appear in some individual terms, it should not be the case when considering the sum of all terms. In section~\ref{sec:Fokker}, we shall see in practice how to compute the particular PN solution.

\subsection{The tail effects at 3PN order in scalar-tensor theories}
\label{sec:tail}

We now focus on the second terms, $\mathcal{H}^{\mu\nu}$ and $\Psi$, in the Eqs.~\eqref{PNsol}, that are the source of the tail effect. They are homogeneous solutions of the wave equation. We follow the algorithm developed in~\cite{Bernard:2017bvn,Marchand:2017pir} to compute the near-zone expansion of homogeneous solutions of the wave equation in $d$ dimensions. The result for $\mathcal{H}^{\mu\nu}$ still stays the same in ST theories. In particular, it starts contributing to the conservative dynamics at $4$PN order. Thus, we only consider the scalar field homogeneous solution $\Psi$.
As we are interested in the $3$PN contribution, it is sufficient to restrict to the quadratic order in the expansion of the scalar field, $\psi=G\psi_{1}+G^{2}\psi_{2}+\mathcal{O}\left(G^{3}\right)$. The equation we want to solve is
\be
\square \psi_{2} = N_{\mathrm{s},2}[h_{1},\psi_{1}]\,,
\ee
where $\square$ is the flat d'Alembertian operator and $N_{\mathrm{s},2}$ is the quadratic part of the source, explicitely given by
\be\label{quadsource}
N_{\mathrm{s},2}[h_{1},\psi_{1}] = \left(1-\frac{2\phi_{0}\omega'_{0}}{d^{2}-d+4\omega_{0}}\right)\eta^{\mu\nu}\p_{\mu}\psi_{1}\p_{\nu}\psi_{1} - h^{\mu\nu}_{1}\p_{\mu\nu}\psi_{1}-\p_{\mu}\psi_{1}\p_{\nu}h^{\mu\nu}_{1}\,,
\ee
where we have also expanded $h$ at quadratic order, $h^{\mu\nu}=G h^{\mu\nu}_{1}+G^{2}h_{2}^{\mu\nu}+\mathcal{O}\left(G^{3}\right)$. We know that the tail effect will result from an interaction between the constant ADM mass $M$ of the system and one time-varying low multipole moment. Thus, we decompose the linearized field as
\sbe
\begin{align}\label{multipoledecomp}
& h_{1}^{\mu\nu} = h_{1,M}^{\mu\nu} + h_{1,{I_{kl}}}^{\mu\nu} \,,\\
& \psi_{1} = \psi_{1,M} + \psi_{1,{I_{j}}} \,,
\end{align}
\see
with
\sbe\label{hpsiIt}
\begin{align}
& h_{1,M}^{00} = -\frac{4}{c^{2}}\tilde{I}\,, &\qquad & h_{1,M}^{0i} = 0\,, &\qquad & h_{1,M}^{ij} = 0 \,, &\qquad\text{and}\quad & \psi_{1,M} = -\frac{2}{c^{2}}\tilde{I}_{\mathrm{s}} \,,\\
& h_{1,{I_{kl}}}^{} = -\frac{2}{c^{2}}\p_{ij}\tilde{I}_{ij}\,, &\qquad & h_{1,{I_{kl}}}^{0i} = \frac{2}{c^{3}}\p_{j}\tilde{I}^{(1)}_{ij}\,, &\qquad & h_{1,{I_{kl}}}^{ij} = -\frac{2}{c^{4}}\tilde{I}_{ij}^{(2)} \,, &\qquad\text{and}\quad & \psi_{1,{I_{j}}} = \frac{2}{c^{2}}\p_{i}\tilde{I}^{i}_{\mathrm{s}} \,,
\end{align}
\see
where
\be
\tilde{I}_{L}(t,r) = \frac{\tilde{k}}{r^{d-2}}\int_{1}^{+\infty}\ud z\,\gamma_{\frac{1-d}{2}}(z)I_{L}\left(t-\frac{zr}{c}\right)\,,
\ee
is the homegeneous retarded solution of the d'Alembertian operator. Note that the lowest time-varying multipole moment in ST theories is the dipole moment, instead of the quadrupole moment in GR. The static mass monopoles are given by
\be
\tilde{I} = \frac{\tilde{k}I}{r^{d-2}}\,, \qquad\text{and}\qquad \tilde{I}_{\mathrm{s}}=\frac{\tilde{k}I_{\mathrm{s}}}{r^{d-2}}\,.
\ee
Inserting the decomposition~\eqref{multipoledecomp} into Eq.~\eqref{quadsource} and keeping only the terms contributing to the tails, we get 
\be
N_{\mathrm{s},2}^{\mathrm{tail}} = 2\left(1-\frac{2\phi_{0}\omega'_{0}}{d^{2}-d+4\omega_{0}}\right)\,\p_{\alpha}\psi_{1,M}\p^{\alpha}\psi_{1,{I_{j}}} - \frac{1}{c^{2}}h^{00}_{1,M}\p_{t}^{2}\psi_{1,{I_{j}}}-h^{\alpha\beta}_{1,{I_{kl}}}\p_{\alpha\beta}\psi_{1,M}\,.
\ee
Using Eqs.~\eqref{hpsiIt}, we see that $N_{\mathrm{s},2}^{\mathrm{tail}}$ admits the decomposition
\be
N_{\mathrm{s},2}^{\mathrm{tail}} = \sum_{l=0}^{+\infty}\hat{n}_{L}N_{2,L}^{\mathrm{s,tail}}\,,
\ee
with
\be
N_{2,L}^{\mathrm{s,tail}} = \sum r^{-k-2\varepsilon}\int_{1}^{+\infty}\ud y\, y^{p}\gamma_{\frac{1-d}{2}}(y)\,F_{L}\left(t-\frac{yr}{c}\right)\,,
\ee
where the function $F_{L}$ is made of products of mass multipole moments. The tail contribution to the scalar field is then given by
\be\label{Psi2tail}
\Psi_{2,\mathrm{tail}} = \sum_{j=0}^{+\infty}\frac{1}{c^{2j}}\Delta^{-j}\hat{x}_{L}f_{2,L}^{(2j)}\,,
\ee
where
\be
\Delta^{-j}\hat{x}_{L} \equiv \frac{\Gamma(\ell+\frac{d}{2})}{\Gamma(\ell+j+\frac{d}{2})}\frac{r^{2j}\hat{x}_{L}}{2^{2j}j!} \,.
\ee
The function $f_{2,L}$ can be factorised into the compact form:
\be
f_{2,L}=\sum \frac{(-)^{\ell+k}\,C_\ell^{p,k}}{2\ell+1+\varepsilon}\,\frac{\Gamma(2\varepsilon-\eta)}{\Gamma(\ell+k-1+2\varepsilon-\eta)}\,\int_0^{+\infty} \ud\tau\,\tau^{-2\varepsilon+\eta}\,F_L^{(\ell+k-1)\mu\nu}(t-\tau)\,,
\ee
where the dimensionless coefficients $C_{\ell}^{p,k}$ are
\be
C_\ell^{p,k} = \int_1^{+\infty} \!\ud y\,y^p\,\gamma_{-1-\frac{\varepsilon}{2}}(y)\, \int_1^{+\infty} \!\ud z\,(y+z)^{\ell+k-2+2\varepsilon-\eta}\,\gamma_{-\ell-1-\frac{\varepsilon}{2}}(z)\,.
\ee
These coefficients have been computed and an analytic closed form expression can be found in the Appendix~D of~\cite{Bernard:2017bvn}. Plugging the formulas into the tail equation~\eqref{Psi2tail}, carefully applying the ``$\varepsilon\eta$'' regularisation procedure and expanding everything at $3$PN order, we obtain the scalar tail,
\be\label{ScalarTail}
\Psi_{2,\mathrm{tail}} = -\frac{8M}{3c^{8}\phi_{0}}x^{i}\int_{0}^{+\infty}\ud\tau\,\left[\ln\left(\frac{c\tau\sqrt{\bar{q}\phi_{0}^{1/2}}}{2\ell_{0}}\right)-\frac{1}{2\varepsilon}+\frac{11}{12}\right]\left(I_{\mathrm{s},i}^{(5)}(t-\tau)-I_{\mathrm{s},i}^{(5)}(t+\tau)\right)\,,
\ee
where $\bar{q}\equiv 4\pi\ue^{\gamma_{\mathrm{E}}}$ and $\ell_{0}$ is the caracteristic length associated to dimensional regularisation. Note the appearance of a pole $1/\varepsilon$. Finally, inserting it into the Fokker action, we obtain the tail part of the action
\be
S_{\mathrm{F}}^{\mathrm{tail}} = \frac{2G^{2}M}{3c^{6}}\left(3+2\omega_{0}\right)\int\ud t\,I_{\mathrm{s},i}(t)\int_{0}^{+\infty}\ud\tau\,\left[\ln\left(\frac{c\tau\sqrt{\bar{q}}}{2\ell_{0}}\right)-\frac{1}{2\varepsilon}-\frac{5}{4(3+2\omega_{0})}+\frac{11}{12}\right]\left(I_{\mathrm{s},i}^{(5)}(t-\tau)-I_{\mathrm{s},i}^{(5)}(t+\tau)\right)\,.
\ee
Performing some integrations by part and using the Hadamard \textit{partie finie} (Pf) notation\footnote{For any regular function $f(t)$ tending towards zero sufficiently rapidly when $t\rightarrow +\infty$, the Hadamard \textit{partie finie} is defined as \[\underset{\tau_{0}}{\mathrm{Pf}}\int\ud t'f(t')\equiv\int_{0}^{+\infty}\ud\tau\ln\left(\frac{\tau}{\tau_{0}}\right)\left[f^{(1)}(t-\tau)-f^{(1)}(t+\tau)\right]\]}, we can rewrite the tail part of the action in a symmetric way,
\begin{align}\label{TailsSym}
S_{\mathrm{F}}^{\mathrm{tail}} &= \frac{2G^{2}M}{3c^{6}}\left(3+2\omega_{0}\right)\int\ud t\,I_{\mathrm{s},i}^{(2)}(t)\int_{0}^{+\infty}\ud\tau\,\left[\ln\left(\frac{c\tau\sqrt{\bar{q}}}{2\ell_{0}}\right)-\frac{1}{2\varepsilon}-\frac{5}{4(3+2\omega_{0})}+\frac{11}{12}\right]\left(I_{\mathrm{s},i}^{(3)}(t-\tau)-I_{\mathrm{s},i}^{(3)}(t+\tau)\right) \\
& = \frac{2G^{2}M}{3c^{6}}\left(3+2\omega_{0}\right)\underset{\tau_{0}}{\mathrm{Pf}}\int\int\frac{\ud t\ud t'}{\vert t-t'\vert}\,I_{\mathrm{s},i}^{(2)}(t)I_{\mathrm{s},i}^{(2)}(t')\,.
\end{align}
where we have defined the constant $\tau_{0}\equiv\frac{2\ell_{0}}{c\sqrt{\bar{q}}}\,\ue^{\frac{1}{2\varepsilon}+\frac{5}{4(3+2\omega_{0})}-\frac{11}{12}}$.

\section{The Fokker Lagrangian in ST theories}\label{sec:Fokker}

\subsection{The ``$n+2$'' method}\label{subsec:n+2method}

We now focus on the particular solution $\left(\overline{h}_{\mathrm{part}},\overline{\psi}_{\mathrm{part}}\right)$. It is obtained by a PN iteration of the field equations. Due to some cancellations between the gravitational and matter parts in the Fokker action, it is sufficient to know the metric at roughly half the order we would have expected. This is the so-called ``$n+2$'' method, that was developped in~\cite{Bernard:2015njp} for general relativity. Here, we generalise this method to scalar-tensor theories where we have one additional degree of freedom. As we are only interested in the dynamics at 3PN order, we do the reasonning for odd PN orders and in $d$ dimensions. We reason by induction and we will see that the proof follows the one of~\cite{Bernard:2015njp}, as the scalar field behaves similarly as $h^{00ii}$.
First, we decompose the metric perturbation as
\be\label{decomp}
\overline{h}^{\mu\nu} \longrightarrow \left\{\begin{array}{l} \overline{h}^{00ii} \equiv 2\frac{(d-2)\overline{h}^{00}+\overline{h}^{ii}}{d-1} \,, \\
\overline{h}^{0i} \,, \\
\overline{h}^{ij} \,.\end{array} \right.
\ee
At leading order in $\left(\bar{h},\overline{\psi}\right)$, the gravitational action reads 
\be\label{SgPN}
S_{g} = \frac{c^{4}\phi_{0}^{\frac{d-1}{2}}}{128\pi G} \int
\!\ud t \!\int \ud^{d}\mathbf{x} \,\biggl[
  \frac{d-1}{2(d-2)}\overline{h}^{00ii}\Box \overline{h}^{00ii} - 4
  \overline{h}^{0i}\Box \overline{h}^{0i} + 2\overline{h}^{ij}\Box
  \overline{h}^{ij} - \frac{2}{d-2}\overline{h}^{ii} \Box \overline{h}^{jj} + 2\left(d(d-1)+4\omega_{0}\right)\overline{\psi}\Box\overline{\psi} +\mathcal(O)\bigl(\bar{h}^3,\overline{\psi}^3\bigr) \biggr] \,,
\ee
while the matter action is given by
\begin{equation}\label{SmPN} 
S_{m} = \sum_A m_A c^{2} \int\!\ud t \biggl[ -1 +
  \frac{v_A^{2}}{2c^{2}} -\frac{1}{4}\overline{h}_A^{00ii}-(1-2s_{A})\overline{\psi}  +
  \frac{v_A^{i}}{c}\,\overline{h}_A^{0i} -
  \frac{v_A^{i}v_A^{j}}{2c^{2}}\,\overline{h}_A^{ij} +
  \frac{v_A^{2}}{2(d-2)c^{2}}\,\overline{h}_A^{ii} +
  \calO\bigl(\overline{h}_A^{2},\,c^{-2}\overline{h}_A,\,c^{-2}\overline{\psi}_A\bigr) \biggr] \,.
\end{equation}
Varying this action with respect to the metric and scalar fields, we can see that the leading order of the PN solution is
\be\label{leadorder}
\left(\bar{h}^{00ii},\bar{h}^{0i},\bar{h}^{ij};\overline{\psi}\right) = \mathcal{O}\left(2,3,4;2\right)\,.
\ee
Consider now a solution of the field equations,
\be
\overline{h}_{n}\equiv\left(\bar{h}_{n}^{00ii},\bar{h}_{n}^{0i},\bar{h}_{n}^{ij};\overline{\psi}_{n}\right) =\mathcal{O}\left(n+1,n+2,n+1;n+1\right)\,,
\ee
where $n$ is an odd number and the orders are included. As we schematicaly have $\frac{\delta S_\text{F}}{\delta \overline{h}} \sim c^{4}\left(\square h-\overline{\Sigma}-\overline{T}\right)$, we have the estimates
\begin{subequations}\label{estimates}
\begin{align}
\frac{\delta S_\text{F}}{\delta \overline{h}^{00ii}}\bigl[ \overline{h}_{n}[\bm{y}_B],\bm{y}_A \bigr] &= \calO\big(n-1\big)\,,\\
\frac{\delta S_\text{F}}{\delta \overline{h}^{0i}}\bigl[ \overline{h}_{n}[\bm{y}_B],\bm{y}_A \bigr] &= \calO\big(n\big)\,,\\
\frac{\delta S_\text{F}}{\delta \overline{h}^{ij}}\bigl[  \overline{h}_{n}[\bm{y}_B], \bm{y}_A \bigr] &= \calO\big(n-1\big)\,,\\
\frac{\delta S_\text{F}}{\delta \overline{\psi}}\bigl[  \overline{h}_{n}[\bm{y}_B], \bm{y}_A \bigr] &= \calO\big(n-1\big)\,.
\end{align}
\end{subequations}
We now define the rest of the complete PN solution by
\be
\left(\bar{h},\bar{\psi}\right) = \bar{h}_{n} + \bar{r}_{n+2}\,,
\ee
with
\begin{equation}\label{remainderestimates}
\overline{r}_{n+2} = (\overline{r}^{00ii}_{n+3},\,\overline{r}^{0i}_{n+4},\,\overline{r}^{ij}_{n+3},\,\overline{r}^{\mathrm{s}}_{n+3}) = \calO(n+3,\,n+4,\,n+3;\,n+3)\,,
\end{equation}
and we expand the Fokker action around the $n$th order PN solution,
\be
\begin{aligned}\label{SFexpand}
S_\text{F}\left[\overline{h}[\bm{y}_B],\,\bm{y}_A\right] =\  & S_\text{F}\left[ \overline{h}_{n}[\bm{y}_B],\,\bm{y}_A\right] + \int \!\ud t \!\int  \ud^{d}\mathbf{x} \,\Biggl[\frac{\delta S_\text{F}}{\delta\overline{h}^{00ii}}\left[ \overline{h}_{n}[\bm{y}_B],\,\bm{y}_A\right]\overline{r}^{00ii}_{n+3} \\
& +\frac{\delta S_\text{F}}{\delta \overline{h}^{0i}}\left[\overline{h}_{n}[\bm{y}_B],\,\bm{y}_A\right]\overline{r}^{0i}_{n+4}+ \frac{\delta S_\text{F}}{\delta \overline{h}^{ij}} \left[\overline{h}_{n}[\bm{y}_B],\,\bm{y}_A\right]\overline{r}^{ij}_{n+3}+ \frac{\delta S_\text{F}}{\delta \overline{\psi}} \left[\overline{h}_{n}[\bm{y}_B],\,\bm{y}_A\right]\overline{r}^{\mathrm{s}}_{n+3}+ \cdots \Biggr]\,,
\end{aligned}
\ee
where the ellipsis stand for quadratic or higher order terms. Inserting the estimates~\eqref{estimates} in Eq.~\eqref{SFexpand}, we have
\begin{equation}\label{nPN}
S_\text{F}\bigl[\overline{h}[\bm{y}_B],\,\bm{y}_A\bigr] =
S_\text{F}\bigl[ \overline{h}_{n}[\bm{y}_B],\,\bm{y}_A\bigr] +
\mathcal{O}\left(2n+2\right) \,.
\end{equation}
The action is thus known at $n$PN order as wanted. Note that the quadratic and higher order terms, generically denoted by the ellipsis in~\eqref{SFexpand}, do not change the result as they contribute to a higher order in the action. The reasonning in the case of $n$ even is very similar. Summarizing our result, the ST ``$n+2$'' method is given by the rule: In order to control the Fokker action at the $n$th PN order, it is sufficient to know the metric at the order
\be\label{resultnp2}
\overline{h}_{n}=\left\{
\begin{aligned}
& \calO\big(n+2,\,n+1,\,n+2;\,n+2\big)\quad \text{included} \qquad \text{when $n$ is even} \,,\\[3pt]
& \calO\big(n+1,\,n+2,\,n+1;\,n+1\big)\quad \text{included} \qquad \text{when $n$ is odd} \,.
\end{aligned}
\right.
\ee

\subsection{Iteration of the post-Newtonian solution}\label{subsec:PNsolution}

We now perform the iteration of the post-Newtonian solution. At 3PN order, according to the ``$n+2$'' method, we need to know the metric at the order $\left(4,5,4;4\right)$. As we will use dimensional regularisation to treat all the divergences, we already define all the quantities in $d$ dimensions. We use the decomposition of the metric given by Eq.~\eqref{decomp}, and define the usual PN potentials 
\begin{subequations}\label{metricpot}
\begin{align} 
& \overline{h}^{00ii} = -\frac{4}{c^{2}}V - \frac{4}{c^{4}} \left[\frac{d-1}{d-2}V^{2}-2\frac{d-3}{d-2}K\right]  +\calO\left(\frac{1}{c^{6}}\right)\,,\\
& \overline{h}^{0i} = - \frac{4}{c^{3}} V_{i} - \frac{4}{c^{5}}\biggl(2\hat{R}_{i} + \frac{d-1}{d-2}V V_i\biggr) +\calO\left(\frac{1}{c^{7}}\right)\,,\\
& \overline{h}^{ij} = - \frac{4}{c^{4}}\biggl(\hat{W}_{ij} -\frac{1}{2} \delta_{ij} \hat{W}\biggr) +\calO\left(\frac{1}{c^{6}}\right) \,,\\
& \overline{\psi} = -\frac{2}{c^{2}}\psi_{(0)} + \frac{2}{c^{4}}\left(1-\frac{2\phi_{0}\omega'_{0}}{d(d-1)+4\omega_{0}}\right)\psi_{(0)}^{2} +\calO\left(\frac{1}{c^{6}}\right)\,,
\end{align}
\end{subequations}
with $\hat{W}=\hat{W}_{ii}$. Each PN potential obeys a flat space-time wave equation, sourced by matter source densities and some lower order PN potentials. They read
{\allowdisplaybreaks
\begin{subequations}\label{defpotentials}
\begin{align}
&\, \Box V = - 4 \pi G\, \sigma \,,\\[3pt]
&\, \Box \psi_{(0)} = 4 \pi G\, \sigma_{\mathrm{s}}\,,\\[3pt]
&\, \Box K = - 4 \pi G\, \sigma\,V\,,\\[3pt]
&\, \Box V_{i} = - 4 \pi G\, \sigma_{i}\,,\\[3pt]
&\, \Box \hat{R}_{i} = -\frac{4\pi G}{d-2}\biggl[\frac{5-d}{2}\, V \sigma_i -\frac{d-1}{2}\, V_i\, \sigma\biggr] -\frac{d-1}{d-2}\,\partial_k V\partial_i V_k -\frac{d(d-1)}{4(d-2)^2}\,\partial_t V \partial_i V +\frac{d(d-1)+4\omega_{0}}{4}\p_{i}\psi_{(0)}\p_{t}\psi_{(0)} \,, \\[3pt]
&\, \Box \hat{W}_{ij} = -4\pi G\biggl(\sigma_{ij} -\delta_{ij}\,\frac{\sigma_{kk}}{d-2}\biggr) -\frac{d-1}{2(d-2)}\partial_i V \partial_j V -\frac{d(d-1)+4\omega_{0}}{2}\partial_i \psi_{(0)} \partial_j \psi_{(0)} \,.
\end{align}
\end{subequations}}\noindent
The gravitational constant $G$ appearing in these equations is linked to the usual Newton constant $G_{\mathrm{N}}$ through the relation
\begin{equation}\label{G}
G = G_\text{N}\,\ell_0^{d-3}\,,
\end{equation}
where $\ell_0$ is the caracteristic length associated to dimensional regularisation. The matter source densities are constructed from the components of the stress-energy tensor for point particles,
\begin{equation}\label{Tmunu}
T^{\mu\nu} = \sum_A \frac{m_A(\phi)}{\sqrt{-g}}\,\frac{v_A^\mu
  v_A^\nu}{\sqrt{-[g_{\rho\sigma}]_A \,v_A^{\rho}v_A^{\sigma}/c^{2}}}
\delta^{(d)}(\mathbf{x}-\bm{y}_A)\,.
\end{equation}
They read~\footnote{These definitions have been modified in the last version of the article in order to match their expressions given in~\ref{sec:matterdensities}. During the calculation, only the latter formulas were used so the final result is unaffected by the new definitions.}
\begin{align}\nn\label{sourcedensity}
& \sigma = 2\left(\frac{1}{\phi_{0}^{\tfrac{d-1}{2}}\varphi^3}\right)\frac{(d-2)T^{00} + T^{ii}}{(d-1)c^2}\,,\qquad\sigma_i =
\left(\frac{1}{\phi_{0}^{\tfrac{d-1}{2}}\varphi^3}\right)\frac{T^{0i}}{c}\,,\qquad\sigma_{ij} = \left(\frac{1}{\phi_{0}^{\tfrac{d-1}{2}}\varphi^3}\right)T^{ij} \,,\\[7pt]
& \sigma_{\mathrm{s}} = - \frac{2\sqrt{-g}}{c^{2}\phi_{0}^{\tfrac{d-1}{2}}\sqrt{d(d-1)+4\omega_{0}}\sqrt{d(d-1)+4\omega(\phi)}}\left(T-2\varphi\frac{\p T}{\p\varphi}\right) \,.
\end{align}
Note that, in addition to the new scalar density, we have slightly changed the definition of the usual densities with respect to the GR result~\cite{Blanchet:2013haa} by adding the scalar field in factor. In the Appendix~\ref{sec:matterdensities}, we give the explicit expressions of the matter source densities as a function of the potentials. Finally, the harmonicity conditions ${\p_{\nu}h^{\mu\nu}=0}$ read
{\allowdisplaybreaks
\begin{subequations}
\label{diffident}
\begin{align}
& \partial_t\biggl\{ \frac{d-1}{2(d-2)} V +\frac{1}{2 c^2}\biggl[\hat W +\left(\frac{d-1}{d-2}\right)^2 V^2 -\frac{2(d-1)(d-3)}{(d-2)^2}\,K\biggr] \biggr\} +\partial_i\biggl\{V_i +\frac{2}{c^2}\biggl[\hat R_i +\frac{d-1}{2(d-2)} V V_i\biggr]  \biggr\} = \mathcal{O}\left(\frac{1}{c^{4}}\right) \,,\\[7pt]
& \partial_t V_i +\partial_j\hat W_{ij} -\frac{1}{2}\,\p_{i}\hat W = \mathcal{O}\left(\frac{1}{c^{2}}\right) \,.
\end{align}
\end{subequations}}\noindent
We emphasize that the gravitational field $\overline{h}$ should only verify the harmonicity conditions~\eqref{diffident} when \textit{on-shell}.

\subsection{Dimensional regularisation}\label{subsec:dimreg}

The computation of the Lagrangian involves non-compact support integrals of the type
\be\label{DivergentInt}
I=\int\mathrm{d}^{3}\mathbf{x}\,F(\mathbf{x})\,,
\ee
where $F(\mathbf{x})$ represents a generic function resulting from the PN iteration of the potentials carried out in the previous section, taken in the limit when $d\rightarrow 3$. The integration of such a function leads to two different types of divergences. First, the ultraviolet divergences result from the point-particle approximation that causes the function $F$ to be singular at the points $\mathbf{y}_{1}$ and $\mathbf{y}_{2}$. Then, the infrared divergences come from the fact that the PN solution diverges at infinity. In the present work, we use dimensional regularisation (DR)~\cite{Blanchet:2003gy} to treat both the infrared and ultraviolet divergences appearing in the integrals of the type~\eqref{DivergentInt}. Following the procedure used in previous works in general relativity, the regularisation scheme will proceed in several steps. First, we perform the integration in $3$ dimensions using Hadamard regularisation (HR)~\cite{Blanchet:2000nu} for both UV and IR divergences. In a second step, we compute the difference between HR and DR in the case of the ultraviolet divergences, resulting in the appearance of a pole. Finally, we add the diffence between HR and DR for infrared divergences. The pole that appears after this step should exactely compensate the one coming from the tail term computed in section~\ref{sec:tail}.

\subsubsection{Ultraviolet divergences}

When $r_{1}\rightarrow 0$, the $3$-dimensional function $F$ admits the following expansion, valid for any $\mathcal{N}\in\mathbb{N}$,
\be\label{devFsing}
F(\mathbf{x})=\sum_{a_0\leq a\leq\mathcal{N}}r_{1}^{a}\,\underset{1\ }{f_{a}}(\mathbf{n}_{1}) + o(r_{1}^{\mathcal{N}}) \,.
\ee
The Hadamard regularisation of the spatial integral~\eqref{DivergentInt} is then given by
\begin{align}\label{intPf}\nonumber
I^{\mathrm{HR}} & \equiv \underset{\ell_{1},\ell_{2}}{\mathrm{Pf}}\int\mathrm{d}^{3}\mathbf{x}\,F(\mathbf{x}) \\
& = \lim_{s\rightarrow 0}\Bigl\{ \int_{\mathcal{S}(s)}\mathrm{d}^{3}\mathbf{x}\,F(\mathbf{x}) +4\pi\sum_{a+3<0}\frac{s^{a+3}}{a+3}\left(\frac{F}{r_{1}^{a}}\right)_{1} +4\pi\ln\left(\frac{s}{\ell_{1}}\right)\left(r_{1}^{3}F\right)_{1} +1\leftrightarrow 2 \Bigr\} \,.
\end{align}
where $\ell_{1}$ and $\ell_{2}$ are two constants of regularisations. The integral on the second line is performed on the domain of integration $\mathcal{S}(s)\equiv\mathbb{R}^{3}\setminus\mathcal{B}(\mathbf{y}_{1},s)\cup\mathcal{B}(\mathbf{y}_{2},s)$, where $\mathcal{B}(\mathbf{y}_{A},s)$ is the sphere centered in $\mathbf{y}_{A}$ of radius $s$. When implementing it in the calculation of the Fokker Lagrangian, we obtain a result that depends on the two constants $\ell_{1}$ and $\ell_{2}$. We now turn on implementing dimensional regularisation. In $d=3+\varepsilon$ spatial dimensions, the expansion~\eqref{devFsing} of the function $F^{(d)}$ becomes,
\be
F^{(d)}(\mathbf{x}) = \underset{q_0\leq q\leq q_{1}}{\sum_{p_0\leq p\leq\mathcal{N}}}r_{1}^{p+q\varepsilon} \underset{1\quad}{f_{p,\,q}^{(\varepsilon)}}(\mathbf{n_{1}}) + o(r_{1}^{\mathcal{N}}) \,.
\ee
We further assume that the function $F^{(d)}$ does not have any pole when $\varepsilon\rightarrow 0$. It implies the following relation between the $d$-dimensional and the $3$-dimensional coefficients,
\be\label{FexpandDdim}
\sum_{q=q_0}^{q_1} \underset{1\qquad}{f_{p,\,q}^{(\varepsilon=0)}}(\mathbf{n_{1}}) = \underset{1\ }{f_{p}}(\mathbf{n_{1}}) \,.
\ee
To obtain the dimensionally regularised version of the integral~\eqref{DivergentInt}, we only need to compute the difference between the $d$-dimensional integral $I^{\mathrm{DR}}\equiv\int\mathrm{d}^{d}\mathbf{x}\,F^{d}(\mathbf{x})$ and the HR integral~\eqref{intPf}, and add this result to the previous one. As when $\varepsilon\rightarrow 0$ the two regularisation procedures give identical results outside the particles' position, these contributions will cancel out in the difference. Thus, we only have to carry-out the calculation locally, i.e. in the vicinity of the particles. Denoting $\mathcal{D}I\equiv I^{\mathrm{DR}}-I^{\mathrm{HR}}$ the difference between the two regularised integrals, we have the formula,
\be\label{DI}
\mathcal{D}I = \frac{1}{\varepsilon}\sum_{q=q_0}^{q_1}\left[\frac{1}{q+1}+\varepsilon\ln \ell_{1}\right]\int\mathrm{d}\Omega_{2+\varepsilon}(\mathbf{n_{1}})\,\underset{1\qquad}{f_{-3,\,q}^{(\varepsilon)}}(\mathbf{n_{1}}) + 1\leftrightarrow 2 +\mathcal{O}(\varepsilon) \,.
\ee
Due to the presence of the pole in Eq.~\eqref{DI}, it is very important to perform the angular integration over the $(d-1)$-dimensional sphere, with volume element $\mathrm{d}\Omega_{2+\varepsilon}(\mathbf{n_{1}})$, up to linear order in $\varepsilon$. Note the presence of the offending value $q=-1$ in the sum over $q$ in Eq.~\eqref{DI}. An important test of our calculation, and in turn of the validity of dimensional regularisation, consists in checking that the spherical angular integrals are always zero for $q=-1$. By construction, the constants $\ell_{1}$ and $\ell_{2}$ will be absent from the final result, i.e. after adding Eq.\eqref{intPf} and Eq.~\eqref{DI}, as these are pure HR constants.

\subsubsection{Infrared divergences}

Next, we carry out the regularisation of the infrared divergences. In $3$ dimensions, the expansion of the function $F$, when $r\rightarrow\infty$, is given by
\begin{equation}\label{FdevInfty}
F(\mathbf{x}) = \sum_{p=-p_0}^{N}\frac{1}{r^p}\,f_p(\mathbf{n})+ o\left(\frac{1}{r^N}\right)\,.
\end{equation}
The regularised value of the integral is then
\begin{equation}\label{IHRdef}
I^\text{HR} = \mathop{\text{{\rm FP}}}_{B=0}\int\ud^{3}\mathbf{x}\,\Bigl(\frac{r}{r_0}\Bigr)^B F(\mathbf{x}) \,,
\end{equation}
where we have introduced the regulator $\left(r/r_{0}\right)^B$, with $B\in\mathbb{C}$ and $r_{0}$ is a regularisation constant. The finite part (FP) at $B=0$ means that we take the zeroth power of $B$ in the Laurent expansion when $B\rightarrow 0$ of the integrand $\left(r/r_{0}\right)^B F(\mathbf{x})$.
Similarly, the $d$-dimensional function  $F^{(d)}$ admits the following expansion near infinity
\begin{equation}\label{Fdevddim}
F^{(d)}(\mathbf{x}) = \sum_{p\geqslant -p_0}\sum_{q=-q_0}^{q_1}\frac{1}{r^{p}}\left(\frac{\ell_0}{r}\right)^{q\varepsilon}f^{(\varepsilon)}_{p,q}(\mathbf{n})\,.
\end{equation}
Assuming that the coefficients $f^{(\varepsilon)}_{p,q}$ admit a well-defined limit when $\varepsilon\rightarrow 0$, which is the case at $3$PN order, we have the following relation,
\begin{equation}\label{relcoeff}
f_{p}(\mathbf{n}) = \sum_{q=-q_0}^{q_1}f^{(\varepsilon=0)}_{p,q}(\mathbf{n})\,.
\end{equation}
The difference between the DR and HR integrals is entirely determined by the coefficients $f^{(\varepsilon)}_{p,q}$ in the expansion at infinity  of the function $F^{(d)}$. At leading order in $\varepsilon\rightarrow 0$, we have
\begin{equation}\label{resdiff}
\mathcal{D}I = \sum_q\left[\frac{1}{(q-1)\varepsilon} - \ln\left(\frac{r_0}{\ell_0}\right)\right]\int\ud\Omega_{2+\varepsilon}\,f^{(\varepsilon)}_{3,q}(\mathbf{n})+ \mathcal{O}\left(\varepsilon\right)\,,
\end{equation}
As for the ultraviolet regularisation procedure, the presence of the pole in Eq.~\eqref{resdiff} implies that the spherical angular integral has to be performed in $d$ dimensions up to linear order in $\varepsilon$. Note also the problematic case $q=1$ in the sum over $q$. During the calculation one should check that the corresponding terms do not appear in our end result.

\subsection{Implementation of the calculation}\label{subsec:implementation}

Once the Fokker Lagrangian has been computed using dimensional regularisation, we can add the Lagrangian describing the tail computed in section~\ref{sec:tail}. We rewrite Eq.~\eqref{TailsSym} by dividing the logarithmic kernel as,
\be\label{Logdecomp}
\ln\left(\frac{\tau}{\tau_{0}}\right)=\ln\left(\frac{c\tau}{2r_{12}}\right)+\ln\left(\frac{2r_{12}}{c\tau_{0}}\right)\,,
\ee
where we recall that $\tau_{0}=\frac{2\ell_{0}}{c\sqrt{\bar{q}}}\,\ue^{\frac{1}{2\varepsilon}+\frac{5}{4(3+2\omega_{0})}-\frac{11}{12}}$, with $\bar{q}=4\pi\ue^{\gamma_{\mathrm{E}}}$. Thanks to this rewriting, one can see that the pole coming from the tails~\eqref{TailsSym} directly cancels the one coming from the dimensional regularisation of the infrared divergences.

Finally, the last step consists in renormalising our result by absorbing the ultraviolet pole through some redefinition of the trajectory of the particles. The complete 3PN shift on the trajectories of the particle that allows to remove the pole $\propto 1/\varepsilon$ is given by,
\begin{align}
\boldsymbol{\delta y}_{\text{3PN}}={}& \frac{\alpha^3 \tilde{G}^3 m_{1}^2 m_{2}}{24c^6 \varepsilon r_{12}^2} \mathbf{n}_{12} \bigl(44
 + 44 \overline{\gamma}
 + 11 \overline{\gamma}^2
 - 4 \overline{\delta}_{1}\bigr) \Bigl(-2
 + 6\varepsilon\ln\bigl(\frac{\sqrt{4\pi\phi_{0}\ue^{\gamma_{\mathrm{E}}}}r'_{1}}{\ell_{0}}\bigr)\Bigr)\,,
\end{align}
where the scalar-tensor PN parameters $\overline{\gamma}$ and $\overline{\delta}_{1}$ are defined in Eqs.~\eqref{1PNparam} and~\eqref{2PNparam}. Following previous works in general relativity, we have introduced the gauge constant $r'_{1}$ and $r'_{2}$ to replace the characteristic length scale $\ell_{0}$, such that the logarithmic dependence in our result only appears through the combination $\ln\left(r_{12}/r'_{1}\right)$ and $\ln\left(r_{12}/r'_{2}\right)$. At the end, our result is thus both IR and UV finite.

\section{Results}\label{sec:results}

\subsection{The 3PN acceleration in scalar-tensor theories}\label{subsec:3PNacc}

The 3PN Lagrangian in harmonic coordinates is a generalised one, meaning that it depends not only on the positions $\mathbf{y}_{A}$ and velocities $\mathbf{v}_{A}$ of the particles, but also on the accelerations $\mathbf{a}_{A}$ and their higher order derivatives.

The accelerations of the particles are obtained by writing the generalised Euler-Lagrange equations, see Eq.~\eqref{GenLagEqs}. Following~\cite{Mirshekari:2013vb}, we express them using a finite number of parameters. We define the scalar-tensor parameters:
\begin{equation}
\begin{aligned}
& \tilde{G} \equiv \frac{G(4+2\omega_{0})}{\phi_{0}(3+2\omega_{0})}\,,\qquad & \zeta \equiv \frac{1}{(4+2\omega_{0})}\,, \\
& \lambda_{1} \equiv \frac{\zeta^{2}}{(1-\zeta)}\left.\frac{\ud\omega}{\ud\varphi}\right\vert_{0}\,,\qquad & \lambda_{2} \equiv \frac{\zeta^{3}}{(1-\zeta)}\left.\frac{\ud^{2}\omega}{\ud\varphi^{2}}\right\vert_{0}\,,\qquad & \lambda_{3} \equiv \frac{\zeta^{4}}{(1-\zeta)}\left.\frac{\ud^{3}\omega}{\ud\varphi^{3}}\right\vert_{0}\,, 
\end{aligned}
\end{equation}
as well as the zeroth and higher order sensitivities,
\begin{equation}
\begin{aligned}
& s_{A} \equiv \left.\frac{\ud\ln m_{A}(\phi)}{\ud\ln\phi}\right\vert_{0}\,,\qquad & s'_{A} \equiv \left.\frac{\ud^{2}\ln m_{A}(\phi)}{\ud\ln\phi^{2}}\right\vert_{0}\,,\qquad & s''_{A} \equiv \left.\frac{\ud^{3}\ln m_{A}(\phi)}{\ud\ln\phi^{3}}\right\vert_{0}\,,\qquad & s'''_{A} \equiv \left.\frac{\ud^{4}\ln m_{A}(\phi)}{\ud\ln\phi^{4}}\right\vert_{0}\,.
\end{aligned}
\end{equation}
At Newtonian order, one additional parameter is sufficient to describe the dynamics,
\be
\alpha\equiv 1-\zeta+\zeta\left(1-2s_{1}\right)\left(1-2s_{2}\right)\,,
\ee
while at 1PN three new parameters were introduced. They all read,
\begin{equation}\label{1PNparam}
\begin{aligned}
& \overline{\gamma} \equiv -\frac{2\zeta}{\alpha}\left(1-2s_{1}\right)\left(1-2s_{2}\right)\,,\\
& \overline{\beta}_{1} \equiv \frac{\zeta}{\alpha^{2}}\left(1-2s_{2}\right)^{2}\left(\lambda_{1}\left(1-2s_{1}\right)+2\zeta s'_{1}\right) \,,\qquad & \overline{\beta}_{2} \equiv \frac{\zeta}{\alpha^{2}}\left(1-2s_{1}\right)^{2}\left(\lambda_{1}\left(1-2s_{2}\right)+2\zeta s'_{2}\right)\,.
\end{aligned}
\end{equation}
Note that they are not all independent, as we have the relation $\alpha(2+\overline{\gamma})=2(1-\zeta)$. Then at 2PN, four new parameters were introduced,
\begin{equation}
\begin{aligned}\label{2PNparam}
& \overline{\delta}_{1} \equiv \frac{\zeta\left(1-\zeta\right)}{\alpha^{2}}\left(1-2s_{1}\right)^{2}\,,\qquad \overline{\delta}_{2} \equiv \frac{\zeta\left(1-\zeta\right)}{\alpha^{2}}\left(1-2s_{2}\right)^{2}\,,\\
& \overline{\chi}_{1} \equiv \frac{\zeta}{\alpha^{3}}\left(1-2s_{2}\right)^{3}\left[\left(\lambda_{2}-4\lambda_{1}^{2}+\zeta\lambda_{1}\right)\left(1-2s_{1}\right)-6\zeta\lambda_{1}s'_{1}+2\zeta^{2}s''_{1}\right] \,,\\
& \overline{\chi}_{2} \equiv \frac{\zeta}{\alpha^{3}}\left(1-2s_{1}\right)^{3}\left[\left(\lambda_{2}-4\lambda_{1}^{2}+\zeta\lambda_{1}\right)\left(1-2s_{2}\right)-6\zeta\lambda_{1}s'_{2}+2\zeta^{2}s''_{2}\right]\,. 
\end{aligned}
\end{equation}
Once again, these parameters are not all independent, as we have the relation $16\overline{\delta}_{1}\overline{\delta}_{2} = \overline{\gamma}^{2}(2+\overline{\gamma})^{2}$. Finally at 3PN order we introduce two new parameters, 
\begin{equation}
\begin{aligned}
& \overline{\kappa}_{1} \equiv \frac{\zeta}{\alpha^{4}}\left(1-2s_{2}\right)^{4}\left[\left(\lambda_{3}-13\lambda_{1}\lambda_{2}+28\lambda_{1}^{3}+\zeta\left(3\lambda_{2}-13\lambda_{1}^{2}\right)+\lambda_{1}\zeta^{2}\right)\left(1-2s_{1}\right) \right.\\
&\qquad\qquad\qquad\qquad\qquad \left.+2\zeta\left(19\lambda_{1}^{2}-4\lambda_{2}-4\lambda_{1}\zeta\right)s'_{1}-12\zeta^{2}\lambda_{1}s''_{1}+2\zeta^{3}s'''_{1}\right] \,,\\
& \overline{\kappa}_{2} \equiv \frac{\zeta}{\alpha^{4}}\left(1-2s_{1}\right)^{4}\left[\left(\lambda_{3}-13\lambda_{1}\lambda_{2}+28\lambda_{1}^{3}+\zeta\left(3\lambda_{2}-13\lambda_{1}^{2}\right)+\lambda_{1}\zeta^{2}\right)\left(1-2s_{2}\right)\right. \\
&\qquad\qquad\qquad\qquad\qquad \left.+2\zeta\left(19\lambda_{1}^{2}-4\lambda_{2}-4\lambda_{1}\zeta\right)s'_{2}-12\zeta^{2}\lambda_{1}s''_{2}+2\zeta^{3}s'''_{2}\right]\,.
\end{aligned}
\end{equation}
We write the full 3PN equations of motion in the following form:
\be
\mathbf{a}_{1} = \mathbf{a}_{1}^{\mathrm{N}} + \mathbf{a}_{1}^{1\mathrm{PN}} + \mathbf{a}_{1}^{2\mathrm{PN}} + \mathbf{a}_{1}^{3\mathrm{PN}} \,.
\ee
The 3PN piece is then decomposed into a local part and a non-local one,
\be
\mathbf{a}_{1}^{3\mathrm{PN}} = \mathbf{a}_{1}^{3\mathrm{PN,\,inst}} + \mathbf{a}_{1}^{3\mathrm{PN,\,tail}}\,,
\ee
and the local part is further split into its increasing power of $\tilde{G}$:
\be
\mathbf{a}_{1}^{3\mathrm{PN},\,\mathrm{inst}} = \tilde{G}\,\mathbf{a}_{1}^{3\mathrm{PN,\,(1)}} + \tilde{G}^{2}\,\mathbf{a}_{1}^{3\mathrm{PN,\,(2)}} + \tilde{G}^{3}\,\mathbf{a}_{1}^{3\mathrm{PN,\,(3)}} + \tilde{G}^{4}\,\mathbf{a}_{1}^{3\mathrm{PN,\,(4)}} \,.
\ee
We have
\begin{subequations}
\begin{align}
\mathbf{a}_{1}^{\mathrm{N}} ={}&- \frac{ \tilde{G}\alpha m_{2}}{r_{12}^2} \mathbf{n}_{12}\,,\\[7pt]
\mathbf{a}^{1\mathrm{PN}}_{1}={}& \frac{\tilde{G}^2 \alpha^2 }{r_{12}^{3}} \mathbf{n}_{12}\biggl[\Bigl(5
 + 2 \overline{\gamma}
 + 2 \overline{\beta}_{2}\Bigr) m_{1} m_{2}
 + 2 \Bigl(2
 + \overline{\gamma}
 + \overline{\beta}_{1}\Bigr) m_{2}^2 \biggr]\nonumber\\
& + \frac{\tilde{G} \alpha m_{2}}{r_{12}^2} \biggl(\mathbf{n}_{12} \biggl[\frac{3}{2} (n_{12} v_2)^2
 + 2 \Bigl(2
 + \overline{\gamma}\Bigr) (v_1 v_2)
 + \Bigl(-1
 -  \overline{\gamma}\Bigr) v_1^{2}
 + \Bigl(-2 -  \overline{\gamma}\Bigr) v_2^{2}\biggr]\nonumber\\
& + \mathbf{v}_{1} \biggl[2 \Bigl(2 + \overline{\gamma}\Bigr) (n_{12} v_1)
 + \Bigl(-3 - 2 \overline{\gamma}\Bigr) (n_{12} v_2)\biggr]
 + \mathbf{v}_{2} \biggl[-2 \Bigl(2
 + \overline{\gamma}\Bigr) (n_{12} v_1)
 + \Bigl(3 + 2 \overline{\gamma}\Bigr) (n_{12} v_2)\biggr]
 \biggr)\,,\\[7pt]
\mathbf{a}^{2\mathrm{PN}}_{1}={}&\frac{\tilde{G}^3 \alpha^3 }{r_{12}^4} \mathbf{n}_{12} \biggl[ m_{1} m_{2}^2\Bigl(\frac{1}{2} \bigl(-69
 - 48 \overline{\gamma}
 - 8 \overline{\gamma}^2\bigr)
 - 4 \bigl(3
 + \overline{\gamma}\bigr) \overline{\beta}_{2}
 + \overline{\beta}_{1} \bigl(-15
 - 4 \overline{\gamma}
 + \frac{24 \overline{\beta}_{2}}{\overline{\gamma}}\bigr)\Bigr)\nonumber\\
&\quad + m_{2}^3 \Bigl(- \frac{9}{4} \bigl(2
 + \overline{\gamma}\bigr)^2
 - 4 \bigl(2
 + \overline{\gamma}\bigr) \overline{\beta}_{1}
 -  \overline{\delta}_{2}
 + 2 \overline{\chi}_{1}\Bigr)
 + m_{1}^2 m_{2} \Bigl(\frac{1}{4} \bigl(-57
 - 44 \overline{\gamma}
 - 9 \overline{\gamma}^2\bigr)
 - 4 \bigl(3
 + \overline{\gamma}\bigr) \overline{\beta}_{2}
 -  \overline{\delta}_{1}
 + 2 \overline{\chi}_{2}\Bigr)\biggr]\nonumber\\
& + \frac{\tilde{G}^2 \alpha^2}{r_{12}^3} \Biggl[\mathbf{v}_{2} \biggl(m_{1} m_{2} \biggl[\Bigl(\frac{1}{4} \bigl(63
 + 40 \overline{\gamma}
 + 2 \overline{\gamma}^2\bigr)
 - 2 \overline{\beta}_{2}
 + 2 \overline{\delta}_{1}\Bigr) (n_{12} v_1)
 + \Bigl(\frac{1}{4} \bigl(-55
 - 40 \overline{\gamma}
 - 2 \overline{\gamma}^2\bigr)
 + 4 \overline{\beta}_{2}
 - 2 \overline{\delta}_{1}\Bigr) (n_{12} v_2)\biggr]\nonumber\\
&\quad + m_{2}^2 \biggl[\Bigl(\frac{1}{2} \bigl(2
 + \overline{\gamma}\bigr)^2
 + 2 \overline{\delta}_{2}\Bigr) (n_{12} v_1)
 + \Bigl(- \frac{1}{2} \bigl(-2
 + \overline{\gamma}\bigr) \bigl(2
 + \overline{\gamma}\bigr)
 + 2 \overline{\beta}_{1}
 - 2 \overline{\delta}_{2}\Bigr) (n_{12} v_2)\biggr]\biggr)\nonumber\\
&\ + \mathbf{v}_{1} \biggl(m_{1} m_{2} \biggl[\Bigl(\frac{1}{4} \bigl(-63
 - 40 \overline{\gamma}
 - 2 \overline{\gamma}^2\bigr)
 + 2 \overline{\beta}_{2}
 - 2 \overline{\delta}_{1}\Bigr) (n_{12} v_1)
 + \Bigl(\frac{1}{4} \bigl(55
 + 40 \overline{\gamma}
 + 2 \overline{\gamma}^2\bigr)
 - 4 \overline{\beta}_{2}
 + 2 \overline{\delta}_{1}\Bigr) (n_{12} v_2)\biggr]\nonumber\\
&\quad + m_{2}^2 \biggl[\Bigl(- \frac{1}{2} \bigl(2
 + \overline{\gamma}\bigr)^2
 - 2 \overline{\delta}_{2}\Bigr) (n_{12} v_1)
 + \Bigl(\frac{1}{2} \bigl(-2 + \overline{\gamma}\bigr) \bigl(2
 + \overline{\gamma}\bigr)
 - 2 \overline{\beta}_{1}
 + 2 \overline{\delta}_{2}\Bigr) (n_{12} v_2)\biggr]\biggr)\nonumber\\
&\ + \mathbf{n}_{12} \biggl(m_{2}^2 \biggl[\Bigl(\frac{1}{2} \bigl(2
 + \overline{\gamma}\bigr)^2
 + 2 \overline{\delta}_{2}\Bigr) (n_{12} v_1)^2 + \Bigl(- \bigl(2
 + \overline{\gamma}\bigr)^2
 - 4 \overline{\delta}_{2}\Bigr) (n_{12} v_1) (n_{12} v_2)\nonumber\\
&\quad + \Bigl(\frac{1}{2} \bigl(-6
 + \overline{\gamma}\bigr) \bigl(2
 + \overline{\gamma}\bigr)
 - 4 \overline{\beta}_{1} + 2 \overline{\delta}_{2}\Bigr) (n_{12} v_2)^2
 - 4 \bigl(2 + \overline{\gamma}\bigr) (v_1 v_2)
 - 2 \overline{\beta}_{1} v_1^{2}
 + 2 \bigl(2
 + \overline{\gamma}\bigr) v_2^{2}\biggr]\nonumber\\
&\quad + m_{1} m_{2} \biggl[\Bigl(\frac{1}{2} \bigl(39
 + 26 \overline{\gamma} + \overline{\gamma}^2\bigr)
 - 4 \overline{\beta}_{2}
 + 2 \overline{\delta}_{1}\Bigr) (n_{12} v_1)^2
 + \bigl(-39
 - 26 \overline{\gamma}
 -  \overline{\gamma}^2
 + 8 \overline{\beta}_{2}
 - 4 \overline{\delta}_{1}\bigr) (n_{12} v_1) (n_{12} v_2)\nonumber\\
&\quad + \Bigl(\frac{1}{2} \bigl(1
 + \overline{\gamma}\bigr) \bigl(17
 + \overline{\gamma}\bigr)
 - 8 \overline{\beta}_{2}
 + 2 \overline{\delta}_{1}\Bigr) (n_{12} v_2)^2
 + \bigl(- \frac{5}{2}
 - 2 \overline{\beta}_{2}\bigr) (v_1 v_2)
 + \Bigl(\frac{1}{4} \bigl(-15 - 8 \overline{\gamma}\bigr)
 -  \overline{\beta}_{2}\Bigr) v_1^{2}
 + \bigl(\frac{5}{4}
 + \overline{\beta}_{2}\bigr) v_2^{2}\biggr]\biggr)\Biggr]\nonumber\\
& + \frac{\tilde{G}\alpha m_{2} }{r_{12}^2} \biggl(\mathbf{v}_{1} \biggl[\frac{3}{2} \bigl(3
 + 2 \overline{\gamma}\bigr) (n_{12} v_2)^3
 + \bigl(1 + \overline{\gamma}\bigr) (n_{12} v_2) v_1^{2}
 + (n_{12} v_2) \Bigl(2 \bigl(2
 + \overline{\gamma}\bigr) (v_1 v_2)
 + \bigl(-5
 - 3 \overline{\gamma}\bigr) v_2^{2}\Bigr)\nonumber\\
&\quad + (n_{12} v_1) \Bigl(-3 \bigl(2
 + \overline{\gamma}\bigr) (n_{12} v_2)^2
 - 2 \bigl(2
 + \overline{\gamma}\bigr) (v_1 v_2)
 + 2 \bigl(2
 + \overline{\gamma}\bigr) v_2^{2}\Bigr)\biggr]\nonumber\\
&\ + \mathbf{v}_{2} \biggl[- \frac{3}{2} \bigl(3
 + 2 \overline{\gamma}\bigr) (n_{12} v_2)^3
 + \bigl(-1
 -  \overline{\gamma}\bigr) (n_{12} v_2) v_1^{2}
 + (n_{12} v_1) \Bigl(3 \bigl(2
 + \overline{\gamma}\bigr) (n_{12} v_2)^2
 + 2 \bigl(2
 + \overline{\gamma}\bigr) (v_1 v_2)
 - 2 \bigl(2 + \overline{\gamma}\bigr) v_2^{2}\Bigr)\nonumber\\
&\quad + (n_{12} v_2) \Bigl(-2 \bigl(2
 + \overline{\gamma}\bigr) (v_1 v_2)
 + \bigl(5
 + 3 \overline{\gamma}\bigr) v_2^{2}\Bigr)\biggr]\nonumber\\
&\ + \mathbf{n}_{12} \biggl[- \frac{15}{8} (n_{12} v_2)^4
 + \bigl(-2
 -  \overline{\gamma}\bigr) (v_1 v_2)^2
 + \frac{3}{2} \bigl(1
 + \overline{\gamma}\bigr) (n_{12} v_2)^2 v_1^{2}
 + 2 \bigl(2
 + \overline{\gamma}\bigr) (v_1 v_2) v_2^{2}\nonumber\\
&\quad + (n_{12} v_2)^2 \Bigl(-3 \bigl(2
 + \overline{\gamma}\bigr) (v_1 v_2)
 + \frac{3}{2} \bigl(3
 + \overline{\gamma}\bigr) v_2^{2}\Bigr)
 + \bigl(-2
 -  \overline{\gamma}\bigr) v_2^{4}\biggr]\biggr)\,.
\end{align}
\end{subequations}
At 2PN order, we recover the result from~\cite{Mirshekari:2013vb}. The instantaneous 3PN terms are then given by
\begin{subequations}
\begin{align}
\mathbf{a}^{3\mathrm{PN,\,(1)}}_{1}={}&\frac{\alpha m_{2}}{r_{12}^2} \Biggl[ \mathbf{v}_{1} \biggl(- \frac{15}{8} \bigl(3
 + 2 \overline{\gamma}\bigr) (n_{12} v_2)^5
 + (n_{12} v_2)^3 \Bigl(-3 \bigl(2
 + \overline{\gamma}\bigr) (v_1 v_2)
 + \frac{3}{2} \bigl(8 + 5 \overline{\gamma}\bigr) v_2^{2}\Bigr)\nonumber\\
&\quad + v_1^{2} \Bigl(- \frac{3}{2} \bigl(1
 + \overline{\gamma}\bigr) (n_{12} v_2)^3
 + \bigl(1
 + \overline{\gamma}\bigr) (n_{12} v_2) v_2^{2}\Bigr)
 + (n_{12} v_2) \Bigl(\bigl(-2 -  \overline{\gamma}\bigr) (v_1 v_2)^2
 + 4 \bigl(2
 + \overline{\gamma}\bigr) (v_1 v_2) v_2^{2}\nonumber\\
&\quad + \bigl(-7
 - 4 \overline{\gamma}\bigr) v_2^{4}\Bigr)
 + (n_{12} v_1) \biggl[\frac{15}{4} \bigl(2
 + \overline{\gamma}\bigr) (n_{12} v_2)^4
 - 2 \bigl(2
 + \overline{\gamma}\bigr) (v_1 v_2) v_2^{2}\nonumber\\
&\quad + (n_{12} v_2)^2 \Bigl(3 \bigl(2
 + \overline{\gamma}\bigr) (v_1 v_2)
 - 6 \bigl(2
 + \overline{\gamma}\bigr) v_2^{2}\Bigr)
 + 2 \bigl(2
 + \overline{\gamma}\bigr) v_2^{4}\biggr]\biggr)\nonumber\\
&\ + \mathbf{v}_{2} \biggl(\frac{15}{8} \bigl(3
 + 2 \overline{\gamma}\bigr) (n_{12} v_2)^5
 + (n_{12} v_2)^3 \Bigl(3 \bigl(2
 + \overline{\gamma}\bigr) (v_1 v_2)
 -  \frac{3}{2} \bigl(8
 + 5 \overline{\gamma}\bigr) v_2^{2}\Bigr)\nonumber\\
&\quad + v_1^{2} \Bigl(\frac{3}{2} \bigl(1
 + \overline{\gamma}\bigr) (n_{12} v_2)^3
 + \bigl(-1
 -  \overline{\gamma}\bigr) (n_{12} v_2) v_2^{2}\Bigr)+ (n_{12} v_1) \biggl[- \frac{15}{4} \bigl(2
 + \overline{\gamma}\bigr) (n_{12} v_2)^4
 + 2 \bigl(2
 + \overline{\gamma}\bigr) (v_1 v_2) v_2^{2}\nonumber\\
&\quad + (n_{12} v_2)^2 \Bigl(-3 \bigl(2
 + \overline{\gamma}\bigr) (v_1 v_2)
 + 6 \bigl(2
 + \overline{\gamma}\bigr) v_2^{2}\Bigr)
 - 2 \bigl(2
 + \overline{\gamma}\bigr) v_2^{4}\biggr]\nonumber\\
&\quad + (n_{12} v_2) \Bigl(\bigl(2
 + \overline{\gamma}\bigr) (v_1 v_2)^2
 - 4 \bigl(2
 + \overline{\gamma}\bigr) (v_1 v_2) v_2^{2}
 + \bigl(7
 + 4 \overline{\gamma}\bigr) v_2^{4}\Bigr)\biggr)\nonumber\\
&\ + \mathbf{n}_{12} \biggl[\frac{35}{16} (n_{12} v_2)^6
 + \bigl(-2
 -  \overline{\gamma}\bigr) (v_1 v_2)^2 v_2^{2}
 + (n_{12} v_2)^4 \Bigl(\frac{15}{4} \bigl(2
 + \overline{\gamma}\bigr) (v_1 v_2)
 -  \frac{15}{8} \bigl(4
 + \overline{\gamma}\bigr) v_2^{2}\Bigr)\nonumber\\
&\quad + v_1^{2} \Bigl(- \frac{15}{8} \bigl(1
 + \overline{\gamma}\bigr) (n_{12} v_2)^4
 + \frac{3}{2} \bigl(1
 + \overline{\gamma}\bigr) (n_{12} v_2)^2 v_2^{2}\Bigr)
 + 2 \bigl(2
 + \overline{\gamma}\bigr) (v_1 v_2) v_2^{4}
 + (n_{12} v_2)^2 \Bigl(\frac{3}{2} \bigl(2
 + \overline{\gamma}\bigr) (v_1 v_2)^2\nonumber\\
& - 6 \bigl(2
 + \overline{\gamma}\bigr) (v_1 v_2) v_2^{2}
 + \frac{3}{2} \bigl(5
 + 2 \overline{\gamma}\bigr) v_2^{4}\Bigr)
 + \bigl(-2
 -  \overline{\gamma}\bigr) v_2^{6}\biggr]\Biggr]\,,\\[7pt]
\mathbf{a}^{3\mathrm{PN,\,(2)}}_{1}={}&\frac{\alpha^2}{r_{12}^3} \Biggl\{\mathbf{v}_{2} \Biggl(m_{1} m_{2} \biggl[\Bigl(\frac{1}{12} \bigl(729
 + 888 \overline{\gamma}
 + 226 \overline{\gamma}^2\bigr)
 - 12 \overline{\beta}_{2}
 + \frac{10}{3} \overline{\delta}_{1}\Bigr) (n_{12} v_1)^3\nonumber\\
&\qquad + \Bigl(\frac{1}{4} \bigl(-565
 - 728 \overline{\gamma}
 - 192 \overline{\gamma}^2\bigr)
 + 32 \overline{\beta}_{2}
 - 8 \overline{\delta}_{1}\Bigr) (n_{12} v_1)^2 (n_{12} v_2)
 + \Bigl(\frac{1}{12} \bigl(95
 - 168 \overline{\gamma}
 - 112 \overline{\gamma}^2\bigr)\nonumber\\
&\qquad + \frac{8}{3} \overline{\delta}_{1}\Bigr) (n_{12} v_2)^3
 + \Bigl(\frac{1}{8} \bigl(137
 + 208 \overline{\gamma}
 + 50 \overline{\gamma}^2\bigr)
 - 10 \overline{\beta}_{2}
 + \overline{\delta}_{1}\Bigr) (n_{12} v_2) v_1^{2}\nonumber\\
&\qquad + (n_{12} v_2) \biggl(\Bigl(\frac{1}{4} \bigl(-27
 - 128 \overline{\gamma}
 - 46 \overline{\gamma}^2\bigr)
 + 12 \overline{\beta}_{2}
 + 2 \overline{\delta}_{1}\Bigr) (v_1 v_2)
 + \Bigl(\frac{1}{8} \bigl(-83
 + 48 \overline{\gamma}
 + 42 \overline{\gamma}^2\bigr)
 - 2 \overline{\beta}_{2}
 - 3 \overline{\delta}_{1}\Bigr) v_2^{2}\biggr)\nonumber\\
&\qquad + (n_{12} v_1) \biggl(\Bigl(\frac{1}{4} \bigl(269
 + 488 \overline{\gamma}
 + 154 \overline{\gamma}^2\bigr)
 - 24 \overline{\beta}_{2}
 + 2 \overline{\delta}_{1}\Bigr) (n_{12} v_2)^2
 + \Bigl(2 \bigl(18
 + 29 \overline{\gamma}
 + 8 \overline{\gamma}^2\bigr)
 - 16 \overline{\beta}_{2}\Bigr) (v_1 v_2)\nonumber\\
&\qquad + \Bigl(\frac{1}{8} \bigl(-207
 - 272 \overline{\gamma}
 - 66 \overline{\gamma}^2\bigr)
 + 9 \overline{\beta}_{2}
 -  \overline{\delta}_{1}\Bigr) v_1^{2}
 + \Bigl(\frac{1}{8} \bigl(-81
 - 192 \overline{\gamma}
 - 62 \overline{\gamma}^2\bigr)
 + 7 \overline{\beta}_{2}
 + \overline{\delta}_{1}\Bigr) v_2^{2}\biggr)\biggr]\nonumber\\
&\quad + m_{2}^2 \biggl[\Bigl(\frac{1}{2} \bigl(2
 + \overline{\gamma}\bigr)^2
 + 2 \overline{\delta}_{2}\Bigr) (n_{12} v_1)^2 (n_{12} v_2)
 + \Bigl(\frac{1}{2} \bigl(2
 + \overline{\gamma}\bigr) \bigl(-2
 + 3 \overline{\gamma}\bigr)
 - 4 \overline{\beta}_{1}
 + 6 \overline{\delta}_{2}\Bigr) (n_{12} v_2)^3\nonumber\\
&\qquad - 2 \overline{\beta}_{1} (n_{12} v_2) v_1^{2}
 + (n_{12} v_2) \biggl(\Bigl(\bigl(-2
 + \overline{\gamma}\bigr) \bigl(2
 + \overline{\gamma}\bigr)
 + 4 \overline{\delta}_{2}\Bigr) (v_1 v_2)
 + \Bigl(- \bigl(-2
 + \overline{\gamma}\bigr) \bigl(2
 + \overline{\gamma}\bigr)
 + 2 \overline{\beta}_{1}
 - 4 \overline{\delta}_{2}\Bigr) v_2^{2}\biggr)\nonumber\\
&\qquad + (n_{12} v_1) \biggl(\Bigl(-2 \bigl(2
+ \overline{\gamma}\bigr)^2
 - 8 \overline{\delta}_{2}\Bigr) (n_{12} v_2)^2
 + \Bigl(- \frac{1}{2} \bigl(2
 + \overline{\gamma}\bigr)^2
 - 2 \overline{\delta}_{2}\Bigr) (v_1 v_2)
 + \Bigl(\frac{1}{2} \bigl(2
 + \overline{\gamma}\bigr)^2
 + 2 \overline{\delta}_{2}\Bigr) v_2^{2}\biggr)\biggr]\Biggr)\nonumber\\
&\ + \mathbf{v}_{1} \Biggl(m_{1} m_{2} \biggl[\Bigl(\frac{1}{12} \bigl(-729
 - 888 \overline{\gamma}
 - 226 \overline{\gamma}^2\bigr)
 + 12 \overline{\beta}_{2}
 -  \frac{10}{3} \overline{\delta}_{1}\Bigr) (n_{12} v_1)^3
 + \Bigl(\frac{1}{4} \bigl(565
 + 728 \overline{\gamma}
 + 192 \overline{\gamma}^2\bigr)\nonumber\\
&\qquad - 32 \overline{\beta}_{2}
 + 8 \overline{\delta}_{1}\Bigr) (n_{12} v_1)^2 (n_{12} v_2)
 + \Bigl(\frac{1}{12} \bigl(-95
 + 168 \overline{\gamma}
 + 112 \overline{\gamma}^2\bigr)
 -  \frac{8}{3} \overline{\delta}_{1}\Bigr) (n_{12} v_2)^3\nonumber\\
&\qquad + \Bigl(\frac{1}{8} \bigl(-137
 - 208 \overline{\gamma}
 - 50 \overline{\gamma}^2\bigr)
  + 10 \overline{\beta}_{2}
 -  \overline{\delta}_{1}\Bigr) (n_{12} v_2) v_1^{2}
 + (n_{12} v_1) \biggl(\Bigl(\frac{1}{4} \bigl(-269
 - 488 \overline{\gamma}
 - 154 \overline{\gamma}^2\bigr)
 + 24 \overline{\beta}_{2}\nonumber\\
&\qquad - 2 \overline{\delta}_{1}\Bigr) (n_{12} v_2)^2
 + \Bigl(-2 \bigl(18
 + 29 \overline{\gamma}
 + 8 \overline{\gamma}^2\bigr)
 + 16 \overline{\beta}_{2}\Bigr) (v_1 v_2)
 + \Bigl(\frac{1}{8} \bigl(207
 + 272 \overline{\gamma}
 + 66 \overline{\gamma}^2\bigr)
 - 9 \overline{\beta}_{2}
 + \overline{\delta}_{1}\Bigr) v_1^{2}\nonumber\\
&\qquad + \Bigl(\frac{1}{8} \bigl(81
 + 192 \overline{\gamma}
 + 62 \overline{\gamma}^2\bigr)
 - 7 \overline{\beta}_{2}
 -  \overline{\delta}_{1}\Bigr) v_2^{2}\biggr)
 + (n_{12} v_2) \biggl(\Bigl(\frac{1}{4} \bigl(27
 + 128 \overline{\gamma}
 + 46 \overline{\gamma}^2\bigr)
 - 12 \overline{\beta}_{2}
 - 2 \overline{\delta}_{1}\Bigr) (v_1 v_2)\nonumber\\
&\qquad + \Bigl(\frac{1}{8} \bigl(83
 - 48 \overline{\gamma}
 - 42 \overline{\gamma}^2\bigr)
 + 2 \overline{\beta}_{2}
 + 3 \overline{\delta}_{1}\Bigr) v_2^{2}\biggr)\biggr]\nonumber\\
&\quad + m_{2}^2 \biggl[\Bigl(- \frac{1}{2} \bigl(2
 + \overline{\gamma}\bigr)^2
 - 2 \overline{\delta}_{2}\Bigr) (n_{12} v_1)^2 (n_{12} v_2)
 + \Bigl(- \frac{1}{2} \bigl(2
 + \overline{\gamma}\bigr) \bigl(-2
 + 3 \overline{\gamma}\bigr)
 + 4 \overline{\beta}_{1}
 - 6 \overline{\delta}_{2}\Bigr) (n_{12} v_2)^3\nonumber\\
&\qquad + 2 \overline{\beta}_{1} (n_{12} v_2) v_1^{2}
 + (n_{12} v_1) \biggl(\Bigl(2 \bigl(2
 + \overline{\gamma}\bigr)^2
 + 8 \overline{\delta}_{2}\Bigr) (n_{12} v_2)^2
 + \Bigl(\frac{1}{2} \bigl(2
 + \overline{\gamma}\bigr)^2
 + 2 \overline{\delta}_{2}\Bigr) (v_1 v_2)\nonumber\\
&\qquad + \Bigl(- \frac{1}{2} \bigl(2
 + \overline{\gamma}\bigr)^2
 - 2 \overline{\delta}_{2}\Bigr) v_2^{2}\biggr)
 + (n_{12} v_2) \biggl(\Bigl(- \bigl(-2
 + \overline{\gamma}\bigr) \bigl(2
 + \overline{\gamma}\bigr)
 - 4 \overline{\delta}_{2}\Bigr) (v_1 v_2)\nonumber\\
&\qquad + \Bigl(\bigl(-2
 + \overline{\gamma}\bigr) \bigl(2
 + \overline{\gamma}\bigr)
 - 2 \overline{\beta}_{1}
 + 4 \overline{\delta}_{2}\Bigr) v_2^{2}\biggr)\biggr]\Biggr)\nonumber\\
&\ + \mathbf{n}_{12} \Biggl[m_{2}^2 \Biggl(\Bigl(- \frac{3}{2} \bigl(2
 + \overline{\gamma}\bigr)^2
 - 6 \overline{\delta}_{2}\Bigr) (n_{12} v_1)^2 (n_{12} v_2)^2
 + \Bigl(- \frac{3}{2} \bigl(-2
 + \overline{\gamma}\bigr) \bigl(2
 + \overline{\gamma}\bigr)
 + 6 \overline{\beta}_{1}
 - 6 \overline{\delta}_{2}\Bigr) (n_{12} v_2)^4\nonumber\\
&\qquad + 2 \bigl(2
 + \overline{\gamma}\bigr) (v_1 v_2)^2
 + 4 \overline{\beta}_{1} (n_{12} v_2)^2 v_1^{2}
 - 4 \bigl(2
 + \overline{\gamma}\bigr) (v_1 v_2) v_2^{2}
 + (n_{12} v_2)^2 \biggl(\Bigl(- \bigl(-6
 + \overline{\gamma}\bigr) \bigl(2
 + \overline{\gamma}\bigr)\nonumber\\
&\qquad - 4 \overline{\delta}_{2}\Bigr) (v_1 v_2)
 + \Bigl(\bigl(-6
 + \overline{\gamma}\bigr) \bigl(2
 + \overline{\gamma}\bigr)
 - 4 \overline{\beta}_{1}
 + 4 \overline{\delta}_{2}\Bigr) v_2^{2}\biggr)
 + (n_{12} v_1) \biggl[\Bigl(3 \bigl(2
 + \overline{\gamma}\bigr)^2
 + 12 \overline{\delta}_{2}\Bigr) (n_{12} v_2)^3\nonumber\\
&\qquad + (n_{12} v_2) \biggl(\Bigl(\bigl(2
 + \overline{\gamma}\bigr)^2
 + 4 \overline{\delta}_{2}\Bigr) (v_1 v_2)
 + \Bigl(- \bigl(2
 + \overline{\gamma}\bigr)^2
 - 4 \overline{\delta}_{2}\Bigr) v_2^{2}\biggr)\biggr]
 + 2 \bigl(2
 + \overline{\gamma}\bigr) v_2^{4}\Biggr)\nonumber\\
&\quad + m_{1} m_{2} \Biggl(\Bigl(\frac{1}{8} \bigl(-171
 - 54 \overline{\gamma}
 - 20 \overline{\gamma}^2\bigr)
 - 18 \overline{\beta}_{2}
 - 10 \overline{\delta}_{1}\Bigr) (n_{12} v_1)^4
 + \Bigl(\frac{1}{2} \bigl(171
 + 54 \overline{\gamma}
 + 20 \overline{\gamma}^2\bigr)
 + 72 \overline{\beta}_{2}\nonumber\\
&\qquad + 40 \overline{\delta}_{1}\Bigr) (n_{12} v_1)^3 (n_{12} v_2)
 + \Bigl(\frac{1}{8} \bigl(-455
 - 294 \overline{\gamma}
 - 32 \overline{\gamma}^2\bigr)
 - 16 \overline{\delta}_{1}\Bigr) (n_{12} v_2)^4
 + \Bigl(\frac{1}{4} \bigl(-177
 - 152 \overline{\gamma}
 - 40 \overline{\gamma}^2\bigr)\nonumber\\
&\qquad - 8 \overline{\beta}_{2}
 - 8 \overline{\delta}_{1}\Bigr) (v_1 v_2)^2
 + \Bigl(\frac{1}{8} \bigl(-91
 - 76 \overline{\gamma}
 - 20 \overline{\gamma}^2\bigr)
 -  \frac{9}{4} \overline{\beta}_{2}
 - 2 \overline{\delta}_{1}\Bigr) v_1^{4}
 + \bigl(43
 + 38 \overline{\gamma}
 + 10 \overline{\gamma}^2\nonumber\\
&\qquad + 7 \overline{\beta}_{2}
 + 8 \overline{\delta}_{1}\bigr) (v_1 v_2) v_2^{2}
 + v_1^{2} \biggl(\Bigl(\frac{1}{4} \bigl(191
 + 124 \overline{\gamma}
 + 33 \overline{\gamma}^2\bigr)
 + 20 \overline{\beta}_{2}
 + 13 \overline{\delta}_{1}\Bigr) (n_{12} v_2)^2\nonumber\\
&\qquad + \Bigl(\frac{1}{2} \bigl(91
 + 76 \overline{\gamma}
 + 20 \overline{\gamma}^2\bigr)
 + 9 \overline{\beta}_{2}
 + 8 \overline{\delta}_{1}\Bigr) (v_1 v_2)
 + \Bigl(\frac{1}{4} \bigl(-91
 - 76 \overline{\gamma}
 - 20 \overline{\gamma}^2\bigr)
 -  \frac{9}{2} \overline{\beta}_{2}
 - 4 \overline{\delta}_{1}\Bigr) v_2^{2}\biggr)\nonumber\\
&\qquad + (n_{12} v_1)^2 \biggl(\Bigl(- \frac{3}{4} \bigl(241
 + 102 \overline{\gamma}
 + 22 \overline{\gamma}^2\bigr)
 - 96 \overline{\beta}_{2}
 - 66 \overline{\delta}_{1}\Bigr) (n_{12} v_2)^2
 + \Bigl(\frac{1}{2} \bigl(-229
 - 176 \overline{\gamma}
 - 49 \overline{\gamma}^2\bigr)\nonumber\\
&\qquad - 36 \overline{\beta}_{2}
 - 26 \overline{\delta}_{1}\Bigr) (v_1 v_2)
 + \Bigl(\frac{1}{4} \bigl(229
 + 176 \overline{\gamma}
 + 49 \overline{\gamma}^2\bigr)
 + 18 \overline{\beta}_{2}
 + 13 \overline{\delta}_{1}\Bigr) v_1^{2}
 + \Bigl(\frac{1}{4} \bigl(229
 + 176 \overline{\gamma}
 + 49 \overline{\gamma}^2\bigr)\nonumber\\
&\qquad + 18 \overline{\beta}_{2}
 + 13 \overline{\delta}_{1}\Bigr) v_2^{2}\biggr)
 + (n_{12} v_2)^2 \biggl(\Bigl(- \frac{5}{2} \bigl(45
 + 32 \overline{\gamma}
 + 7 \overline{\gamma}^2\bigr)
 - 24 \overline{\beta}_{2}
 - 30 \overline{\delta}_{1}\Bigr) (v_1 v_2)\nonumber\\
&\qquad + \Bigl(\frac{1}{4} \bigl(259
 + 196 \overline{\gamma}
 + 37 \overline{\gamma}^2\bigr)
 + 4 \overline{\beta}_{2}
 + 17 \overline{\delta}_{1}\Bigr) v_2^{2}\biggr)
 + (n_{12} v_1) \biggl[\Bigl(\frac{1}{2} \bigl(383
 + 198 \overline{\gamma}
 + 26 \overline{\gamma}^2\bigr)\nonumber\\
&\qquad + 48 \overline{\beta}_{2}
 + 52 \overline{\delta}_{1}\Bigr) (n_{12} v_2)^3
 + \Bigl(\frac{1}{2} \bigl(-205
 - 148 \overline{\gamma}
 - 41 \overline{\gamma}^2\bigr)
 - 36 \overline{\beta}_{2}
 - 26 \overline{\delta}_{1}\Bigr) (n_{12} v_2) v_1^{2}\nonumber\\
&\qquad + (n_{12} v_2) \biggl(\Bigl(2 \bigl(122
 + 87 \overline{\gamma}
 + 21 \overline{\gamma}^2\bigr)
 + 64 \overline{\beta}_{2}
 + 56 \overline{\delta}_{1}\Bigr) (v_1 v_2)
 + \Bigl(\frac{1}{2} \bigl(-283
 - 200 \overline{\gamma}
 - 43 \overline{\gamma}^2\bigr)\nonumber\\
&\qquad - 28 \overline{\beta}_{2}
 - 30 \overline{\delta}_{1}\Bigr) v_2^{2}\biggr)\biggr]
 + \Bigl(\frac{1}{8} \bigl(-81
 - 76 \overline{\gamma}
 - 20 \overline{\gamma}^2\bigr)
 -  \frac{5}{4} \overline{\beta}_{2}
 - 2 \overline{\delta}_{1}\Bigr) v_2^{4}\Biggr)\Biggr]\Biggr\}\,,\\[7pt]
\mathbf{a}^{3\mathrm{PN,\,(3)}}_{1}={}&\frac{\alpha^2 m_{1}^2 m_{2}}{r_{12}^4} \biggl[\mathbf{v}_{1} \biggl(\Bigl(- \frac{11}{4} \overline{\gamma} \bigl(2
 + \overline{\gamma}\bigr)
 + \frac{\bigl(10 + \overline{\gamma}\bigr) \overline{\delta}_{1}}{2 + \overline{\gamma}}\Bigr) (n_{12} v_1)
 + \Bigl(\frac{11}{4} \overline{\gamma} \bigl(2
 + \overline{\gamma}\bigr)
 -  \frac{\bigl(10 + \overline{\gamma}\bigr) \overline{\delta}_{1}}{2 + \overline{\gamma}}\Bigr) (n_{12} v_2)\biggr)\nonumber\\
&\ + \mathbf{v}_{2} \biggl(\Bigl(\frac{11}{4} \overline{\gamma} \bigl(2
 + \overline{\gamma}\bigr)
 -  \frac{\bigl(10 + \overline{\gamma}\bigr) \overline{\delta}_{1}}{2 + \overline{\gamma}}\Bigr) (n_{12} v_1)
 + \Bigl(- \frac{11}{4} \overline{\gamma} \bigl(2
 + \overline{\gamma}\bigr)
 + \frac{\bigl(10 + \overline{\gamma}\bigr) \overline{\delta}_{1}}{2 + \overline{\gamma}}\Bigr) (n_{12} v_2)\biggr)\nonumber\\
&\ + \mathbf{n}_{12} \biggl(\Bigl(\frac{55}{8} \overline{\gamma} \bigl(2
 + \overline{\gamma}\bigr)
 -  \frac{5 \bigl(10 + \overline{\gamma}\bigr) \overline{\delta}_{1}}{2 \bigl(2 + \overline{\gamma}\bigr)}\Bigr) (n_{12} v_1)^2
 + \Bigl(- \frac{55}{4} \overline{\gamma} \bigl(2
 + \overline{\gamma}\bigr)
 + \frac{5 \bigl(10 + \overline{\gamma}\bigr) \overline{\delta}_{1}}{2 + \overline{\gamma}}\Bigr) (n_{12} v_1) (n_{12} v_2)\nonumber\\
&\qquad + \Bigl(\frac{55}{8} \overline{\gamma} \bigl(2
 + \overline{\gamma}\bigr)
 -  \frac{5 \bigl(10 + \overline{\gamma}\bigr) \overline{\delta}_{1}}{2 \bigl(2 + \overline{\gamma}\bigr)}\Bigr) (n_{12} v_2)^2
 + \Bigl(\frac{11}{4} \overline{\gamma} \bigl(2
 + \overline{\gamma}\bigr)
 -  \frac{\bigl(10 + \overline{\gamma}\bigr) \overline{\delta}_{1}}{2 + \overline{\gamma}}\Bigr) (v_1 v_2)\nonumber\\
&\qquad + \Bigl(- \frac{11}{8} \overline{\gamma} \bigl(2
 + \overline{\gamma}\bigr)
 + \frac{\bigl(10 + \overline{\gamma}\bigr) \overline{\delta}_{1}}{2 \bigl(2 + \overline{\gamma}\bigr)}\Bigr) v_1^{2}
 + \Bigl(- \frac{11}{8} \overline{\gamma} \bigl(2
 + \overline{\gamma}\bigr)
 + \frac{\bigl(10 + \overline{\gamma}\bigr) \overline{\delta}_{1}}{2 \bigl(2 + \overline{\gamma}\bigr)}\Bigr) v_2^{2}\biggr)\biggr]\nonumber\\
& + \frac{\alpha^3}{r_{12}^4} \Biggl[\mathbf{v}_{1} \Biggl(m_{1} m_{2}^2 \biggl[\biggl(\frac{1}{24} \bigl(-921
 - 1040 \overline{\gamma}
 - 234 \overline{\gamma}^2
 + 24 \overline{\gamma}^3\bigr) + \frac{2}{3} \bigl(35
 + 9 \overline{\gamma}\bigr) \overline{\delta}_{1}
 + \overline{\beta}_{1} \Bigl(\frac{1}{2} \bigl(65
 + 44 \overline{\gamma}\bigr)\nonumber\\
&\qquad + \frac{24 \overline{\beta}_{2}}{\overline{\gamma}}
 -  \frac{24 \overline{\delta}_{1}}{\overline{\gamma}}\Bigr)
 + \frac{1}{3} \bigl(53
 + 18 \overline{\gamma}\bigr) \overline{\delta}_{2} -  \frac{24 \overline{\beta}_{2} \overline{\delta}_{2}}{\overline{\gamma}}
 + \pi^2 \Bigl(- \frac{3}{128} \bigl(2
 + \overline{\gamma}\bigr) \bigl(-82
 - 34 \overline{\gamma}
 + 7 \overline{\gamma}^2\bigr)\nonumber\\
&\qquad -  \frac{21}{64} \bigl(2
 + \overline{\gamma}\bigr) \overline{\delta}_{1}
 -  \frac{21}{64} \bigl(2 + \overline{\gamma}\bigr) \overline{\delta}_{2}\Bigr)\biggr) (n_{12} v_1)
 + \biggl(\frac{1}{24} \bigl(1437
 + 1232 \overline{\gamma}
 + 222 \overline{\gamma}^2
 - 24 \overline{\gamma}^3\bigr)
 -  \frac{2}{3} \bigl(35
 + 9 \overline{\gamma}\bigr) \overline{\delta}_{1}\nonumber\\
&\qquad + \overline{\beta}_{1} \Bigl(\frac{1}{2} \bigl(-43
 - 44 \overline{\gamma}\bigr)
 -  \frac{48 \overline{\beta}_{2}}{\overline{\gamma}}
 + \frac{24 \overline{\delta}_{1}}{\overline{\gamma}}\Bigr)
 + \frac{1}{3} \bigl(-59
 - 18 \overline{\gamma}\bigr) \overline{\delta}_{2}
 + \overline{\beta}_{2} \Bigl(4 \bigl(3
 + \overline{\gamma}\bigr) + \frac{24 \overline{\delta}_{2}}{\overline{\gamma}}\Bigr)\nonumber\\
&\qquad + \pi^2 \Bigl(\frac{3}{128} \bigl(2
 + \overline{\gamma}\bigr) \bigl(-82
 - 34 \overline{\gamma}
 + 7 \overline{\gamma}^2\bigr)
 + \frac{21}{64} \bigl(2
 + \overline{\gamma}\bigr) \overline{\delta}_{1}
 + \frac{21}{64} \bigl(2 + \overline{\gamma}\bigr) \overline{\delta}_{2}\Bigr)\biggr) (n_{12} v_2)\biggr]\nonumber\\
&\quad + m_{2}^3 \biggl(\Bigl(\frac{1}{2} \bigl(2
 + \overline{\gamma}\bigr)^3
 + 2 \bigl(2
 + \overline{\gamma}\bigr) \overline{\delta}_{2}\Bigr) (n_{12} v_1)
 + \Bigl(- \frac{1}{4} \bigl(2 + \overline{\gamma}\bigr)^2 \bigl(-5
 + 2 \overline{\gamma}\bigr)
 + 4 \bigl(2
 + \overline{\gamma}\bigr) \overline{\beta}_{1}\nonumber\\
&\qquad + \bigl(-3
 - 2 \overline{\gamma}\bigr) \overline{\delta}_{2}
 - 2 \overline{\chi}_{1}\Bigr) (n_{12} v_2)\biggr)\nonumber\\
&\quad + m_{1}^2 m_{2} \biggl[\biggl(\frac{1}{12} \bigl(1325
 + 1328 \overline{\gamma}
 + 411 \overline{\gamma}^2
 + 39 \overline{\gamma}^3\bigr)
 + \frac{1}{2} \bigl(63
 + 40 \overline{\gamma}\bigr) \overline{\beta}_{2}
 + \bigl(5
 + \overline{\gamma}\bigr) \overline{\delta}_{1} + 4 \overline{\chi}_{2}\nonumber\\
&\qquad + \Bigl(\frac{33}{2} \bigl(2
 + \overline{\gamma}\bigr)^2
 - 6 \overline{\delta}_{1}\Bigr) \ln\bigl(r'_{1}\bigr)
 + \Bigl(- \frac{33}{2} \bigl(2
 + \overline{\gamma}\bigr)^2
 + 6 \overline{\delta}_{1}\Bigr) \ln\bigl(r_{12}\bigr)\biggr) (n_{12} v_1)\nonumber\\
&\qquad + \biggl(\frac{1}{12} \bigl(-1463
 - 1484 \overline{\gamma}
 - 438 \overline{\gamma}^2
 - 39 \overline{\gamma}^3\bigr)
 + \frac{1}{2} \bigl(-43
 - 40 \overline{\gamma}\bigr) \overline{\beta}_{2}
 + \bigl(-6
 -  \overline{\gamma}\bigr) \overline{\delta}_{1} - 6 \overline{\chi}_{2}\nonumber\\
&\qquad + \Bigl(- \frac{33}{2} \bigl(2
 + \overline{\gamma}\bigr)^2
 + 6 \overline{\delta}_{1}\Bigr) \ln\bigl(r'_{1}\bigr)
 + \Bigl(\frac{33}{2} \bigl(2
 + \overline{\gamma}\bigr)^2 - 6 \overline{\delta}_{1}\Bigr) \ln\bigl(r_{12}\bigr)\biggr) (n_{12} v_2)\biggr]\Biggr)\nonumber\\
&\  + \mathbf{v}_{2} \Biggl(m_{1} m_{2}^2 \biggl[\biggl(\frac{1}{24} \bigl(921
 + 1040 \overline{\gamma}
 + 234 \overline{\gamma}^2
 - 24 \overline{\gamma}^3\bigr) -  \frac{2}{3} \bigl(35
 + 9 \overline{\gamma}\bigr) \overline{\delta}_{1}
 + \overline{\beta}_{1} \Bigl(\frac{1}{2} \bigl(-65
 - 44 \overline{\gamma}\bigr)
 -  \frac{24 \overline{\beta}_{2}}{\overline{\gamma}}
 + \frac{24 \overline{\delta}_{1}}{\overline{\gamma}}\Bigr)\nonumber\\
&\qquad + \frac{1}{3} \bigl(-53
 - 18 \overline{\gamma}\bigr) \overline{\delta}_{2} + \frac{24 \overline{\beta}_{2} \overline{\delta}_{2}}{\overline{\gamma}}
 + \pi^2 \Bigl(\frac{3}{128} \bigl(2
 + \overline{\gamma}\bigr) \bigl(-82
 - 34 \overline{\gamma}
 + 7 \overline{\gamma}^2\bigr)
 + \frac{21}{64} \bigl(2
 + \overline{\gamma}\bigr) \overline{\delta}_{1}\nonumber\\
&\qquad + \frac{21}{64} \bigl(2 + \overline{\gamma}\bigr) \overline{\delta}_{2}\Bigr)\biggr) (n_{12} v_1)
 + \biggl(\frac{1}{24} \bigl(-1437
 - 1232 \overline{\gamma}
 - 222 \overline{\gamma}^2
 + 24 \overline{\gamma}^3\bigr)
 + \frac{2}{3} \bigl(35
 + 9 \overline{\gamma}\bigr) \overline{\delta}_{1}\nonumber\\
&\qquad + \overline{\beta}_{1} \Bigl(\frac{1}{2} \bigl(43
 + 44 \overline{\gamma}\bigr)
 + \frac{48 \overline{\beta}_{2}}{\overline{\gamma}}
 -  \frac{24 \overline{\delta}_{1}}{\overline{\gamma}}\Bigr)
 + \frac{1}{3} \bigl(59
 + 18 \overline{\gamma}\bigr) \overline{\delta}_{2}
 + \overline{\beta}_{2} \Bigl(-4 \bigl(3
 + \overline{\gamma}\bigr) -  \frac{24 \overline{\delta}_{2}}{\overline{\gamma}}\Bigr)\nonumber\\
&\qquad + \pi^2 \Bigl(- \frac{3}{128} \bigl(2
 + \overline{\gamma}\bigr) \bigl(-82
 - 34 \overline{\gamma}
 + 7 \overline{\gamma}^2\bigr)
 -  \frac{21}{64} \bigl(2
 + \overline{\gamma}\bigr) \overline{\delta}_{1}
 -  \frac{21}{64} \bigl(2 + \overline{\gamma}\bigr) \overline{\delta}_{2}\Bigr)\biggr) (n_{12} v_2)\biggr]\nonumber\\
&\quad + m_{2}^3 \biggl(\Bigl(- \frac{1}{2} \bigl(2
 + \overline{\gamma}\bigr)^3
 - 2 \bigl(2
 + \overline{\gamma}\bigr) \overline{\delta}_{2}\Bigr) (n_{12} v_1)
 + \Bigl(\frac{1}{4} \bigl(2 + \overline{\gamma}\bigr)^2 \bigl(-5
 + 2 \overline{\gamma}\bigr)
 - 4 \bigl(2
 + \overline{\gamma}\bigr) \overline{\beta}_{1}\nonumber\\
&\qquad + \bigl(3
 + 2 \overline{\gamma}\bigr) \overline{\delta}_{2}
 + 2 \overline{\chi}_{1}\Bigr) (n_{12} v_2)\biggr)\nonumber\\
&\quad + m_{1}^2 m_{2} \biggl[\biggl(\frac{1}{12} \bigl(-1325
 - 1328 \overline{\gamma}
 - 411 \overline{\gamma}^2
 - 39 \overline{\gamma}^3\bigr)
 + \frac{1}{2} \bigl(-63
 - 40 \overline{\gamma}\bigr) \overline{\beta}_{2}
 + \bigl(-5 -  \overline{\gamma}\bigr) \overline{\delta}_{1}
 - 4 \overline{\chi}_{2}\nonumber\\
&\qquad + \Bigl(- \frac{33}{2} \bigl(2
 + \overline{\gamma}\bigr)^2
 + 6 \overline{\delta}_{1}\Bigr) \ln\bigl(r'_{1}\bigr)
 + \Bigl(\frac{33}{2} \bigl(2
 + \overline{\gamma}\bigr)^2 - 6 \overline{\delta}_{1}\Bigr) \ln\bigl(r_{12}\bigr)\biggr) (n_{12} v_1)\nonumber\\
&\qquad + \biggl(\frac{1}{12} \bigl(1463
 + 1484 \overline{\gamma}
 + 438 \overline{\gamma}^2
 + 39 \overline{\gamma}^3\bigr)
 + \frac{1}{2} \bigl(43 + 40 \overline{\gamma}\bigr) \overline{\beta}_{2}
 + \bigl(6
 + \overline{\gamma}\bigr) \overline{\delta}_{1}
 + 6 \overline{\chi}_{2}\nonumber\\
&\qquad + \Bigl(\frac{33}{2} \bigl(2
 + \overline{\gamma}\bigr)^2
 - 6 \overline{\delta}_{1}\Bigr) \ln\bigl(r'_{1}\bigr)
 + \Bigl(- \frac{33}{2} \bigl(2
 + \overline{\gamma}\bigr)^2 + 6 \overline{\delta}_{1}\Bigr) \ln\bigl(r_{12}\bigr)\biggr) (n_{12} v_2)\biggr]\Biggr)\nonumber\\
&\  + \mathbf{n}_{12} \Biggl(m_{2}^3 \biggl(\Bigl(- \frac{1}{4} \bigl(1
 + \overline{\gamma}\bigr) \bigl(2
 + \overline{\gamma}\bigr)^2
 + \bigl(-1 -  \overline{\gamma}\bigr) \overline{\delta}_{2}\Bigr) (n_{12} v_1)^2
 + \Bigl(\frac{1}{2} \bigl(1
 + \overline{\gamma}\bigr) \bigl(2
 + \overline{\gamma}\bigr)^2\nonumber\\
&\qquad + 2 \bigl(1
 + \overline{\gamma}\bigr) \overline{\delta}_{2}\Bigr) (n_{12} v_1) (n_{12} v_2) + \Bigl(- \frac{1}{8} \bigl(2
 + \overline{\gamma}\bigr)^2 \bigl(-43
 + 2 \overline{\gamma}\bigr)
 + 10 \bigl(2
 + \overline{\gamma}\bigr) \overline{\beta}_{1}
 + \frac{1}{2} \bigl(3
 - 2 \overline{\gamma}\bigr) \overline{\delta}_{2}
 - 5 \overline{\chi}_{1}\Bigr) (n_{12} v_2)^2\nonumber\\
&\qquad + \Bigl(\frac{9}{2} \bigl(2
 + \overline{\gamma}\bigr)^2
 + 8 \bigl(2
 + \overline{\gamma}\bigr) \overline{\beta}_{1}
 + 2 \overline{\delta}_{2}\Bigr) (v_1 v_2)
 - 2 \overline{\chi}_{1} v_1^{2}
 + \Bigl(- \frac{9}{4} \bigl(2
 + \overline{\gamma}\bigr)^2
 - 4 \bigl(2 + \overline{\gamma}\bigr) \overline{\beta}_{1}
 -  \overline{\delta}_{2}\Bigr) v_2^{2}\biggr)\nonumber\\
&\quad + m_{1} m_{2}^2 \biggl[\biggl(\frac{1}{24} \bigl(1245
 + 878 \overline{\gamma}
 - 264 \overline{\gamma}^2
 - 192 \overline{\gamma}^3\bigr)
 + \frac{1}{3} \bigl(-181 - 18 \overline{\gamma}\bigr) \overline{\delta}_{1}
 + \overline{\beta}_{1} \Bigl(\frac{1}{4} \bigl(-299
 - 192 \overline{\gamma}\bigr)\nonumber\\
&\qquad -  \frac{40 \overline{\beta}_{2}}{\overline{\gamma}}
 + \frac{60 \overline{\delta}_{1}}{\overline{\gamma}}\Bigr)
 + \frac{1}{3} \bigl(-83
 - 18 \overline{\gamma}\bigr) \overline{\delta}_{2} + \overline{\beta}_{2} \Bigl(-8 \bigl(2
 + \overline{\gamma}\bigr)
 + \frac{60 \overline{\delta}_{2}}{\overline{\gamma}}\Bigr)\nonumber\\
&\qquad + \pi^2 \Bigl(\frac{15}{256} \bigl(2
 + \overline{\gamma}\bigr) \bigl(-82
 - 34 \overline{\gamma}
 + 7 \overline{\gamma}^2\bigr)
 + \frac{105}{128} \bigl(2 + \overline{\gamma}\bigr) \overline{\delta}_{1}
 + \frac{105}{128} \bigl(2
 + \overline{\gamma}\bigr) \overline{\delta}_{2}\Bigr)\biggr) (n_{12} v_1)^2\nonumber\\
&\qquad + \biggl(\frac{1}{12} \bigl(-1125
 - 806 \overline{\gamma}
 + 270 \overline{\gamma}^2
 + 192 \overline{\gamma}^3\bigr) + \frac{2}{3} \bigl(181
 + 18 \overline{\gamma}\bigr) \overline{\delta}_{1}
 + \overline{\beta}_{1} \Bigl(\frac{1}{2} \bigl(299
 + 192 \overline{\gamma}\bigr)
 + \frac{80 \overline{\beta}_{2}}{\overline{\gamma}}
 -  \frac{120 \overline{\delta}_{1}}{\overline{\gamma}}\Bigr)\nonumber\\
&\qquad + \frac{4}{3} \bigl(43
 + 9 \overline{\gamma}\bigr) \overline{\delta}_{2} + \overline{\beta}_{2} \Bigl(16 \bigl(2
 + \overline{\gamma}\bigr)
 -  \frac{120 \overline{\delta}_{2}}{\overline{\gamma}}\Bigr)
 + \pi^2 \Bigl(- \frac{15}{128} \bigl(2
 + \overline{\gamma}\bigr) \bigl(-82
 - 34 \overline{\gamma}
 + 7 \overline{\gamma}^2\bigr)\nonumber\\
&\qquad -  \frac{105}{64} \bigl(2 + \overline{\gamma}\bigr) \overline{\delta}_{1}
 -  \frac{105}{64} \bigl(2
 + \overline{\gamma}\bigr) \overline{\delta}_{2}\Bigr)\biggr) (n_{12} v_1) (n_{12} v_2)
 + \biggl(\frac{1}{24} \bigl(3339
 + 2294 \overline{\gamma}
 - 36 \overline{\gamma}^2 - 192 \overline{\gamma}^3\bigr)\nonumber\\
&\qquad + \frac{1}{3} \bigl(-181
 - 18 \overline{\gamma}\bigr) \overline{\delta}_{1}
 + \overline{\beta}_{1} \Bigl(\frac{1}{4} \bigl(-129
 - 152 \overline{\gamma}\bigr)
 -  \frac{100 \overline{\beta}_{2}}{\overline{\gamma}}
 + \frac{60 \overline{\delta}_{1}}{\overline{\gamma}}\Bigr) + \frac{1}{3} \bigl(-89
 - 18 \overline{\gamma}\bigr) \overline{\delta}_{2}\nonumber\\
&\qquad + \overline{\beta}_{2} \Bigl(2 \bigl(7
 + \overline{\gamma}\bigr)
 + \frac{60 \overline{\delta}_{2}}{\overline{\gamma}}\Bigr)
 + \pi^2 \Bigl(\frac{15}{256} \bigl(2
 + \overline{\gamma}\bigr) \bigl(-82
 - 34 \overline{\gamma} + 7 \overline{\gamma}^2\bigr)
 + \frac{105}{128} \bigl(2
 + \overline{\gamma}\bigr) \overline{\delta}_{1}\nonumber\\
&\qquad + \frac{105}{128} \bigl(2
 + \overline{\gamma}\bigr) \overline{\delta}_{2}\Bigr)\biggr) (n_{12} v_2)^2
 + \biggl(\frac{1}{6} \bigl(198
 + 17 \overline{\gamma} - 117 \overline{\gamma}^2
 - 36 \overline{\gamma}^3\bigr)
 -  \frac{70}{3} \overline{\delta}_{1}
 + \overline{\beta}_{1} \Bigl(- \frac{5}{2} \bigl(5
 + 4 \overline{\gamma}\bigr)\nonumber\\
&\qquad -  \frac{8 \overline{\beta}_{2}}{\overline{\gamma}}
 + \frac{24 \overline{\delta}_{1}}{\overline{\gamma}}\Bigr)
 -  \frac{20}{3} \overline{\delta}_{2} + \overline{\beta}_{2} \Bigl(4 \bigl(1
 + \overline{\gamma}\bigr)
 + \frac{24 \overline{\delta}_{2}}{\overline{\gamma}}\Bigr)
 + \pi^2 \Bigl(\frac{3}{128} \bigl(2
 + \overline{\gamma}\bigr) \bigl(-82
 - 34 \overline{\gamma}
 + 7 \overline{\gamma}^2\bigr)\nonumber\\
&\qquad + \frac{21}{64} \bigl(2
 + \overline{\gamma}\bigr) \overline{\delta}_{1} + \frac{21}{64} \bigl(2
 + \overline{\gamma}\bigr) \overline{\delta}_{2}\Bigr)\biggr) (v_1 v_2)
 + \biggl(\frac{1}{12} \bigl(216
 + 271 \overline{\gamma}
 + 165 \overline{\gamma}^2
 + 36 \overline{\gamma}^3\bigr)
 + \frac{35}{3} \overline{\delta}_{1} + \overline{\beta}_{1} \Bigl(\frac{1}{4} \bigl(85
 + 36 \overline{\gamma}\bigr)\nonumber\\
&\qquad -  \frac{20 \overline{\beta}_{2}}{\overline{\gamma}}
 -  \frac{12 \overline{\delta}_{1}}{\overline{\gamma}}\Bigr)
 + \frac{10}{3} \overline{\delta}_{2}
 + \overline{\beta}_{2} \Bigl(2 \bigl(5
 + \overline{\gamma}\bigr)
 -  \frac{12 \overline{\delta}_{2}}{\overline{\gamma}}\Bigr) + \pi^2 \Bigl(- \frac{3}{256} \bigl(2
 + \overline{\gamma}\bigr) \bigl(-82
 - 34 \overline{\gamma}
 + 7 \overline{\gamma}^2\bigr)\nonumber\\
&\qquad -  \frac{21}{128} \bigl(2
 + \overline{\gamma}\bigr) \overline{\delta}_{1}
 -  \frac{21}{128} \bigl(2 + \overline{\gamma}\bigr) \overline{\delta}_{2}\Bigr)\biggr) v_1^{2}
 + \biggl(\frac{1}{12} \bigl(-198
 - 17 \overline{\gamma}
 + 117 \overline{\gamma}^2
 + 36 \overline{\gamma}^3\bigr)\nonumber\\
&\qquad + \frac{35}{3} \overline{\delta}_{1}
 + \overline{\beta}_{1} \Bigl(\frac{5}{4} \bigl(5
 + 4 \overline{\gamma}\bigr) + \frac{4 \overline{\beta}_{2}}{\overline{\gamma}}
 -  \frac{12 \overline{\delta}_{1}}{\overline{\gamma}}\Bigr)
 + \frac{10}{3} \overline{\delta}_{2}
 + \overline{\beta}_{2} \Bigl(-2 \bigl(1
 + \overline{\gamma}\bigr)
 -  \frac{12 \overline{\delta}_{2}}{\overline{\gamma}}\Bigr)\nonumber\\
&\qquad + \pi^2 \Bigl(- \frac{3}{256} \bigl(2
 + \overline{\gamma}\bigr) \bigl(-82 - 34 \overline{\gamma}
 + 7 \overline{\gamma}^2\bigr)
 -  \frac{21}{128} \bigl(2
 + \overline{\gamma}\bigr) \overline{\delta}_{1}
 -  \frac{21}{128} \bigl(2
 + \overline{\gamma}\bigr) \overline{\delta}_{2}\Bigr)\biggr) v_2^{2}\biggr]\nonumber\\
&\quad + m_{1}^2 m_{2} \biggl[\biggl(\frac{1}{24} \bigl(-8959 - 9568 \overline{\gamma}
 - 2865 \overline{\gamma}^2
 - 171 \overline{\gamma}^3\bigr)
 + \frac{1}{4} \bigl(-187
 - 144 \overline{\gamma}\bigr) \overline{\beta}_{2}
 + \frac{1}{2} \bigl(-23
 + 3 \overline{\gamma}\bigr) \overline{\delta}_{1}\nonumber\\
&\qquad - 10 \overline{\chi}_{2} + \Bigl(- \frac{165}{4} \bigl(2
 + \overline{\gamma}\bigr)^2
 + 15 \overline{\delta}_{1}\Bigr) \ln\bigl(r'_{1}\bigr)
 + \Bigl(\frac{165}{4} \bigl(2
 + \overline{\gamma}\bigr)^2
 - 15 \overline{\delta}_{1}\Bigr) \ln\bigl(r_{12}\bigr)\biggr) (n_{12} v_1)^2\nonumber\\
&\qquad + \biggl(\frac{1}{12} \bigl(9268
 + 9760 \overline{\gamma}
 + 2871 \overline{\gamma}^2
 + 171 \overline{\gamma}^3\bigr)
 + \frac{1}{2} \bigl(179
 + 144 \overline{\gamma}\bigr) \overline{\beta}_{2}
 + \bigl(25
 - 3 \overline{\gamma}\bigr) \overline{\delta}_{1} + 20 \overline{\chi}_{2}\nonumber\\
&\qquad + \Bigl(\frac{165}{2} \bigl(2
 + \overline{\gamma}\bigr)^2
 - 30 \overline{\delta}_{1}\Bigr) \ln\bigl(r'_{1}\bigr)
 + \Bigl(- \frac{165}{2} \bigl(2
 + \overline{\gamma}\bigr)^2 + 30 \overline{\delta}_{1}\Bigr) \ln\bigl(r_{12}\bigr)\biggr) (n_{12} v_1) (n_{12} v_2)\nonumber\\
&\qquad + \biggl(\frac{1}{24} \bigl(-8386
 - 9148 \overline{\gamma}
 - 2742 \overline{\gamma}^2
 - 171 \overline{\gamma}^3\bigr) + \frac{1}{4} \bigl(-31
 - 104 \overline{\gamma}\bigr) \overline{\beta}_{2}
 + \frac{1}{2} \bigl(-22
 + 3 \overline{\gamma}\bigr) \overline{\delta}_{1}
 - 15 \overline{\chi}_{2}\nonumber\\
&\qquad + \Bigl(- \frac{165}{4} \bigl(2
 + \overline{\gamma}\bigr)^2 + 15 \overline{\delta}_{1}\Bigr) \ln\bigl(r'_{1}\bigr)
 + \Bigl(\frac{165}{4} \bigl(2
 + \overline{\gamma}\bigr)^2
 - 15 \overline{\delta}_{1}\Bigr) \ln\bigl(r_{12}\bigr)\biggr) (n_{12} v_2)^2\nonumber\\
&\qquad + \biggl(\frac{1}{12} \bigl(-1463 - 1634 \overline{\gamma}
 - 513 \overline{\gamma}^2
 - 33 \overline{\gamma}^3\bigr)
 + \frac{1}{2} \bigl(5
 - 8 \overline{\gamma}\bigr) \overline{\beta}_{2}
 + \bigl(-7
 + \overline{\gamma}\bigr) \overline{\delta}_{1}
 - 4 \overline{\chi}_{2}\nonumber\\
&\qquad + \Bigl(- \frac{33}{2} \bigl(2 + \overline{\gamma}\bigr)^2
 + 6 \overline{\delta}_{1}\Bigr) \ln\bigl(r'_{1}\bigr)
 + \Bigl(\frac{33}{2} \bigl(2
 + \overline{\gamma}\bigr)^2
 - 6 \overline{\delta}_{1}\Bigr) \ln\bigl(r_{12}\bigr)\biggr) (v_1 v_2)
 + \biggl(\frac{1}{24} \bigl(1805 + 1898 \overline{\gamma}\nonumber\\
&\qquad + 567 \overline{\gamma}^2
 + 33 \overline{\gamma}^3\bigr)
 + \frac{1}{4} \bigl(43
 + 24 \overline{\gamma}\bigr) \overline{\beta}_{2}
 + \frac{1}{2} \bigl(9
 -  \overline{\gamma}\bigr) \overline{\delta}_{1}
 + \Bigl(\frac{33}{4} \bigl(2
 + \overline{\gamma}\bigr)^2 - 3 \overline{\delta}_{1}\Bigr) \ln\bigl(r'_{1}\bigr)\nonumber\\
&\qquad + \Bigl(- \frac{33}{4} \bigl(2
 + \overline{\gamma}\bigr)^2
 + 3 \overline{\delta}_{1}\Bigr) \ln\bigl(r_{12}\bigr)\biggr) v_1^{2}
 + \biggl(\frac{1}{24} \bigl(1463
 + 1634 \overline{\gamma} + 513 \overline{\gamma}^2
 + 33 \overline{\gamma}^3\bigr)
 + \frac{1}{4} \bigl(-5
 + 8 \overline{\gamma}\bigr) \overline{\beta}_{2}\nonumber\\
&\qquad + \frac{1}{2} \bigl(7
 -  \overline{\gamma}\bigr) \overline{\delta}_{1}
 + 2 \overline{\chi}_{2}
 + \Bigl(\frac{33}{4} \bigl(2
 + \overline{\gamma}\bigr)^2 - 3 \overline{\delta}_{1}\Bigr) \ln\bigl(r'_{1}\bigr)
 + \Bigl(- \frac{33}{4} \bigl(2
 + \overline{\gamma}\bigr)^2
 + 3 \overline{\delta}_{1}\Bigr) \ln\bigl(r_{12}\bigr)\biggr) v_2^{2}\biggr]\Biggr)\Biggr]\,,\\[7pt]
\mathbf{a}^{3\mathrm{PN,\,(4)}}_{1}={}& \frac{\alpha^3 \mathbf{n}_{12}}{r_{12}^5} \biggl[\Bigl(\frac{11}{12} \overline{\gamma} \bigl(2
 + \overline{\gamma}\bigr)
 -  \frac{\bigl(10 + \overline{\gamma}\bigr) \overline{\delta}_{1}}{3 \bigl(2 + \overline{\gamma}\bigr)}\Bigr) m_{1}^3 m_{2}
 + \Bigl(\frac{11}{2} \overline{\gamma} \bigl(2
 + \overline{\gamma}\bigr)
 -  \frac{2 \overline{\gamma} \overline{\delta}_{1}}{2 + \overline{\gamma}}\Bigr) m_{1}^2 m_{2}^2\nonumber\\
&\quad + \Bigl(\frac{55}{12} \overline{\gamma} \bigl(2
 + \overline{\gamma}\bigr)
 -  \frac{5 \bigl(-2 + \overline{\gamma}\bigr) \overline{\delta}_{2}}{3 \bigl(2 + \overline{\gamma}\bigr)}\Bigr) m_{1} m_{2}^3\biggr] \nonumber\\
& + \frac{\alpha^4 \mathbf{n}_{12}}{r_{12}^5} \biggl(m_{2}^4 \biggl[\frac{8}{3} \bigl(2
 + \overline{\gamma}\bigr) \overline{\delta}_{2}  + \overline{\beta}_{1} \Bigl(\frac{14}{3} \bigl(2
 + \overline{\gamma}\bigr)^2
 + \frac{8}{3} \overline{\delta}_{2}\Bigr)
 + \frac{2}{3} \bigl(24
 + 36 \overline{\gamma}
 + 18 \overline{\gamma}^2
 + 3 \overline{\gamma}^3
 + 2 \overline{\kappa}_{1}\bigr)
 - 4 \bigl(2
 + \overline{\gamma}\bigr) \overline{\chi}_{1}\biggr]\nonumber\\
&\quad + m_{1}^2 m_{2}^2 \biggl[\frac{1}{36} \bigl(6168
 + 5240 \overline{\gamma}
 + 1085 \overline{\gamma}^2
 - 27 \overline{\gamma}^3\bigr)
 + 8 \bigl(\overline{\beta}_{2}\bigr)^2
 + \frac{11}{9} \bigl(-5
 + 3 \overline{\gamma}\bigr) \overline{\delta}_{1} - 4 \overline{\delta}_{2}\nonumber\\
&\qquad + \overline{\beta}_{2} \Bigl(2 \bigl(65
 + 34 \overline{\gamma}
 + 4 \overline{\gamma}^2\bigr)
 + \frac{16 \overline{\delta}_{2}}{\overline{\gamma}}\Bigr)
 + \pi^2 \Bigl(\frac{1}{64} \bigl(2
 + \overline{\gamma}\bigr) \bigl(-82
 - 34 \overline{\gamma}
 + 7 \overline{\gamma}^2\bigr) + \frac{7}{32} \bigl(2
 + \overline{\gamma}\bigr) \overline{\delta}_{1}
 + \frac{7}{32} \bigl(2
 + \overline{\gamma}\bigr) \overline{\delta}_{2}\Bigr)\nonumber\\
&\qquad - 4 \bigl(4
 + \overline{\gamma}\bigr) \overline{\chi}_{2}
 + \overline{\beta}_{1} \Bigl(\frac{1}{3} \bigl(149
 + 80 \overline{\gamma}
 + 14 \overline{\gamma}^2\bigr) -  \frac{4 \bigl(44 + 9 \overline{\gamma}\bigr) \overline{\beta}_{2}}{\overline{\gamma}}
 + \frac{64 \bigl(\overline{\beta}_{2}\bigr)^2}{\overline{\gamma}^2}
 + \frac{8 \bigl(2 + \overline{\gamma}\bigr) \overline{\delta}_{1}}{3 \overline{\gamma}}
 + \frac{32 \overline{\chi}_{2}}{\overline{\gamma}}\Bigr)\biggr]\nonumber\\
&\quad + m_{1}^3 m_{2} \biggl[4 \bigl(\overline{\beta}_{2}\bigr)^2 + \frac{1}{3} \bigl(16
 + 9 \overline{\gamma}\bigr) \overline{\delta}_{1}
 + \overline{\beta}_{2} \Bigl(\frac{1}{3} \bigl(119
 + 74 \overline{\gamma}
 + 14 \overline{\gamma}^2\bigr)
 + \frac{8}{3} \overline{\delta}_{1}\Bigr)
 + \frac{1}{36} \bigl(-563
 - 614 \overline{\gamma} - 72 \overline{\gamma}^2\nonumber\\
&\qquad + 39 \overline{\gamma}^3
 + 48 \overline{\kappa}_{2}\bigr)
 -  \frac{2}{3} \bigl(19
 + 6 \overline{\gamma}\bigr) \overline{\chi}_{2}
 + \Bigl(- \frac{11}{2} \bigl(2
 + \overline{\gamma}\bigr)^2
 + 2 \overline{\delta}_{1}\Bigr) \ln\bigl(r'_{1}\bigr) + \Bigl(\frac{11}{2} \bigl(2
 + \overline{\gamma}\bigr)^2
 - 2 \overline{\delta}_{1}\Bigr) \ln\bigl(r_{12}\bigr)\biggr]\nonumber\\
&\quad + m_{1} m_{2}^3 \biggl[\frac{1}{36} \bigl(6668
 + 6514 \overline{\gamma}
 + 1619 \overline{\gamma}^2
 + 6 \overline{\gamma}^3\bigr) + \bigl(\overline{\beta}_{1}\bigr)^2 \bigl(10
 + \frac{64 \overline{\beta}_{2}}{\overline{\gamma}^2}\bigr)
 -  \frac{86}{9} \overline{\delta}_{1}
 + \overline{\beta}_{1} \Bigl(117
 + 66 \overline{\gamma}
 + 8 \overline{\gamma}^2\nonumber\\
&\qquad -  \frac{2 \bigl(68 + 21 \overline{\gamma}\bigr) \overline{\beta}_{2}}{\overline{\gamma}} + \frac{16 \overline{\delta}_{1}}{\overline{\gamma}}\Bigr)
 + \frac{5}{9} \bigl(5
 + 6 \overline{\gamma}\bigr) \overline{\delta}_{2}
 + \pi^2 \Bigl(\frac{1}{64} \bigl(2
 + \overline{\gamma}\bigr) \bigl(-82
 - 34 \overline{\gamma}
 + 7 \overline{\gamma}^2\bigr)
 + \frac{7}{32} \bigl(2
 + \overline{\gamma}\bigr) \overline{\delta}_{1}\nonumber\\
&\qquad + \frac{7}{32} \bigl(2
 + \overline{\gamma}\bigr) \overline{\delta}_{2}\Bigr)
 - 2 \bigl(9
 + 2 \overline{\gamma}\bigr) \overline{\chi}_{1}
 + \overline{\beta}_{2} \Bigl(\frac{1}{3} \bigl(89
 + 68 \overline{\gamma}
 + 14 \overline{\gamma}^2\bigr)
 + \frac{8 \bigl(2 + \overline{\gamma}\bigr) \overline{\delta}_{2}}{3 \overline{\gamma}} + \frac{32 \overline{\chi}_{1}}{\overline{\gamma}}\Bigr)\nonumber\\
&\qquad + \Bigl(\frac{11}{2} \bigl(2
 + \overline{\gamma}\bigr)^2
 - 2 \overline{\delta}_{2}\Bigr) \ln\bigl(r'_{2}\bigr)
 + \Bigl(- \frac{11}{2} \bigl(2
 + \overline{\gamma}\bigr)^2
 + 2 \overline{\delta}_{2}\Bigr) \ln\bigl(r_{12}\bigr)\biggr]\biggr)\,.
\end{align}
\end{subequations}
Finally, the nonlocal part of the acceleration is given by
\begin{align}\nn
a_{1}^{i\,3\mathrm{PN,\,tail}} = & -\frac{4G^{2}M}{3c^{6}\phi_{0}}(1-2s_{1})\int_{0}^{+\infty}\ud\tau\,\ln\left(\frac{c\tau}{2r_{12}}\right)\left[I_{s\,i}^{(5)}(t-\tau)-I_{s\,i}^{(5)}(t+\tau)\right]\\
& +\frac{8G^{2}M}{3c^{6}\phi_{0}}(1-2s_{1})\left(\left[\ln r_{12}I_{(s)\,i}^{(2)}\right]^{(2)}-\ln r_{12}I_{s\,i}^{(4)}\right) -\frac{4G^{2}M}{3c^{6}m_{1}}(3+2\omega_{0})\frac{n^{i}_{12}}{r_{12}}\left(I_{s\,i}^{(2)}\right)^{2}\,,
\end{align}
where $M=m_{1}+m_{2}$ is the ADM mass. The instantaneous terms on the second line come from the introduction of the time-varying scale $r_{12}$ in the decomposition~\eqref{Logdecomp}. The term on the first line is the nonlocal tail term. Replacing the scalar dipole moment by its explicit expression,
\be
I_{\mathrm{s}}^{i}(t) = -\frac{1}{\phi_{0}\left(3+2\omega_{0}\right)}\left[m_{1}\left(1-2s_{1}\right)y_{1}^{i}+m_{2}\left(1-2s_{2}\right)y_{2}^{i}\right]\,,
\ee
and using the ST parameters to express the instantaneous terms, we get
\begin{align}\nn
a_{1}^{i\,3\mathrm{PN,\,tail}} = & -\frac{4G^{2}M}{3c^{6}\phi_{0}}m_{1}(1-2s_{1})\int_{0}^{+\infty}\ud\tau\,\ln\left(\frac{c\tau}{2r_{12}}\right)\left[I_{s\,i}^{(5)}(t-\tau)-I_{s\,i}^{(5)}(t+\tau)\right]\\
& +\frac{8\tilde{G}^{3}\alpha^{3}M m_{1}^{2}m_{2}}{3c^{6}r_{12}^{4}}\left(\overline{\delta}_{1}+\frac{\overline{\gamma}(2+\overline{\gamma})}{4}\right)\left[2(n_{12}v_{12})v_{12}^{i}-8(n_{12}v_{12})^{2}n_{12}^{i}+v_{12}^{2}n_{12}^{i}-\frac{\tilde{G}\alpha M}{r_{12}}n_{12}^{i}\right] \\
& -\frac{4\tilde{G}^{4}\alpha^{4}M m_{1}^{2}m_{2}^{2}}{3c^{6}r_{12}^{5}}\left(\overline{\delta}_{1}+\frac{\overline{\gamma}(2+\overline{\gamma})}{2}+\overline{\delta}_{2}\right)n^{i}_{12}\,.
\end{align}

\subsection{Discussions}\label{subsec:discuss}

\paragraph*{\textbf{General comments.}}

We have verified that our result is manifestly Lorentz invariant, as it is expected because we are in harmonic coordinates and dimensional regularisation does not break the Lorentz-Poincar\'e symmetry. Then in the GR limit, i.e. when $\omega_{0}\rightarrow \infty$ and $\phi_{0}\rightarrow 1$, we recover the 3PN acceleration of GR, up to an unphysical shift of the trajectories of the particles. The presence of such a shift only reflects the freedom we have when performing the redefinition of the trajectories of the particles in order to remove the pole.
Finally, up to 2PN, the equations of motion depend only on the constant $\alpha$ through the combination $\tilde{G}\alpha m_{A}$. At 3PN, this is no more the case and an additional dependence on $\alpha$ appears in some terms. This is a new and unexpected result. One way of seeing it is to rewrite it as a dependence on $\zeta$, through the relation ${1-\zeta = \alpha\left(1+\overline{\gamma}/2\right)}$. It is then clear that it introduces an explicit dependence on the function $\omega_{0}$. However, depending on the compact objects we are considering such a particularity may disappear, and thus it may be difficult to see the observational consequence of such a dependence.

\paragraph*{\textbf{The binary black hole limit.}}

An important test of our result consists in studying the binary black hole limit. We have seen that the sensitivity of a stationary black hole is exactly given by $s=1/2$. If we assume that $s_{A}=1/2$ still holds for each black hole in a binary system, our result is indistinguishible from GR, up to a simple rescalling of the mass. In particular, the nonlocal tail part of the acceleration does not contribute and the explicit dependence in $\zeta$ disappears. This result confirms that Hawking's theorem may hold also for binary black holes, which is \textit{a priori} not a stationary system, at least up to 3PN order. However, the 3PN dynamics only describes the early-inspiral phase of the coalescence. In particular, it does not tell us anything about the late-inspiral phase where strong-field effects appear and Hawking's theorem may break down. A correct implementation of such hypothetical effects can only be done using the ST EOB formalism coupled to full numerical relativity results for ST theories. Some numerical results~\cite{Healy:2011ef} have shown that, unless an external mechanism activates the dynamics of the scalar field, binary black holes in ST theories and GR are indistinguishable.

\paragraph*{\textbf{Black hole -- neutron star binary.}}

We now consider the case when one of the compact object is a black hole, say $s_{1}=1/2$, while the other one is a neutron star, with $s_{2}\approx 0.2$. First, we find that the explicit dependence in $\zeta$ also disappears for this configuration, up to an unphysical shift. Then, as we have $\overline{\gamma}=\overline{\delta}_{1}=\overline{\beta}_{i}=\overline{\kappa}_{i}=\overline{\chi}_{i}=0$, the final result depends only on one single parameter,
\be
\overline{\delta}_{2}=\frac{\zeta}{1-\zeta}\left(1-2s_{2}\right)^{2}\,.
\ee
It means that the 3PN equations differs from GR only through this only parameter. Thus, if this result still holds for the gravitational waveforms\footnote{It has already been shown that it is the case for the tensor gravitational waveform~\cite{Lang:2014osa}.}, the black hole -- neutron star system may not allow to distinguish between Brans-Dicke theory (with constant function $\omega$), and general scalar-tensor theories. Of course, this conclusion does not apply when dynamical scalarisation takes place~\cite{Barausse:2012da,Palenzuela:2013hsa}, a situation that is not described by our prescription for the matter through a skeletonized action~\cite{Sennett:2016rwa}.

\paragraph*{\textbf{Concluding remarks.}}

In the companion paper~\cite{Bernard:2018}, we compute the conserved integrals of motion and the reduction to the center-of-mass frame. Due to the presence of the non-local term in the action~\eqref{TailsSym}, the computation of the conserved energy and angular momentum has to be treated carefully, as some extra contributions may appear~\cite{Bernard:2016wrg}.

Finally, in scalar-tensor theories, the finite-size effects are expected to start contributing to the dynamics at 3PN order~\cite{Esposito-Farese:2011cha}\footnote{The tidal effects may even start at a lower order (1PN) due to some dynamical scalarisation phenomenon that could be responsible for the large value of some coefficients in the expansion of the mass w.r.t. the scalar field~\cite{Esposito-Farese:2011cha} .}. They may prove very usefull to constrain the theory as such effects can have a different signature in the signal. Thus, if we want to capture the full gravitational waveform  at 2PN order in ST theories, the tidal effects should be properly included in the 3PN dynamics. As it is a work on its own, we have not considered these effects in this paper, and have left it for a future work.

\acknowledgments 

The author thanks L. Blanchet, G. Faye, A. Heffernan and C. Will for usefull discussions and comments, and a carefull reading of the manuscript. She also thanks Félix-Louis Julié and David Trestini for pointing out some typos in the published article.
The author acknowledges financial support provided under the European Union's H2020 ERC Consolidator Grant ``Matter and strong-field gravity: New frontiers in Einstein's theory'' grant agreement no. MaGRaTh-646597. The author thankfully acknowledges the computer resources, technical expertise and assistance provided by CENTRA/IST. Computations were performed at the cluster “Baltasar-Sete-S\' ois” and supported by the H2020 ERC Consolidator Grant "Matter and strong field gravity: New frontiers in Einstein's theory" grant agreement no. MaGRaTh-646597."


\appendix

\section{Demonstration of Eq.~(3.2)}\label{sec:Lemma1}

In this appendix, we give the proof of Eq.~\eqref{lemma1} in the case of dimensional regularisation. It mainly follows the proof done for Hadamard regularisation in~\cite{Bernard:2015njp}. We consider the difference
\be
\Delta_{\mathrm{g}}=L_{\mathrm{g}}-\int\ud^{d}x\,\mathcal{M}\left(\mathcal{L}_{\mathrm{g}}\right)\,,
\ee
where $L_{\mathrm{g}}=\int\ud^{3}x\,\mathcal{L}_{\mathrm{g}}$ involves only the complete solution. As it is perfectly regular everywhere, we don't need any regularisation. Thus, we can add a regulator in the integral without altering the result,
\be\label{Deltag}
\Delta_{\mathrm{g}}=\int\ud^{d}x\,\left[\mathcal{L}_{\mathrm{g}}-\mathcal{M}\left(\mathcal{L}_{\mathrm{g}}\right)\right]\,.
\ee
Now, as the complete solution $\mathcal{L}_{\mathrm{g}}$ coincide with $\mathcal{M}\left(\mathcal{L}_{\mathrm{g}}\right)$ outside the source, the integrand of~\eqref{Deltag} is zero in the exterior region. Thus, it is of compact support around the source and we can PN expand Eq.~\eqref{Deltag} without changing the result,
\be\label{DeltagPN}
\Delta_{\mathrm{g}}=\int\ud^{d}x\,\left[\overline{\mathcal{L}_{\mathrm{g}}}-\overline{\mathcal{M}\left(\mathcal{L}_{\mathrm{g}}\right)}\right]\,.
\ee
Then, the matching equation~\eqref{MatchingEq} implies a common structure of the Lagrangian densities, namely
\be\label{Lagdensstructure}
\overline{\mathcal{M}\left(\mathcal{L}_{\mathrm{g}}\right)}=\mathcal{M}\left(\overline{\mathcal{L}}_{\mathrm{g}}\right)\sim\sum\hat{n}_{L}r^{a}(\ln r)^{b}F(t)\,,
\ee
where $a\in\mathbb{Z}$, $b\in\mathbb{N}$, and the functions $F(t)$ are functions of the source multipole moments. Inserting Eq.~\eqref{Lagdensstructure} into the integral involving $\overline{\mathcal{M}\left(\mathcal{L}_{\mathrm{g}}\right)}$ in Eq.~\eqref{DeltagPN}, one can see that it involves integrals of the type $\int\ud^{d}x\,\hat{n}_{L}r^{a}(\ln r)^{b}F(t)$. After performing the angular integration, one is left with the simple radial integrals, $\int\ud r\,r^{a+2+\varepsilon}(\ln r)^{b}$, where we have written the dimension $d=3+\varepsilon$. These integrals are all zero by analytic continuation in $\varepsilon\in\mathbb{C}$. To show this, we split this integral into a near-zone integral, $\int_{r<\mathcal{R}}$, and a far-zone integral, $\int_{r>\mathcal{R}}$. The near-zone integral is computed for $\mathrm{Re}(\varepsilon)>-a-3$, and analytically continued for $\varepsilon\in\mathbb{C}$, except for the value $\varepsilon=-a-3$. Similarly the far-zone integral is computed for $\mathrm{Re}(\varepsilon)<-a-3$, and analytically continued for $\varepsilon\in\mathbb{C}$, except for the value $\varepsilon=-a-3$. Then, summing the two analytic continuations, one find that they cancel each other and the total integral is zero for any $\varepsilon\in\mathbb{C}$. Finally, one gets that $\int\ud^{d}x\,\overline{\mathcal{M}\left(\mathcal{L}_{\mathrm{g}}\right)}=0$, and as a consequence,
\be
\Delta_{\mathrm{g}}=\int\ud^{d}x\,\overline{\mathcal{L}_{\mathrm{g}}}\,.
\ee
This ends our proof.

\section{The matter source densities in scalar-tensor theories }\label{sec:matterdensities}

In this appendix, we write the explicit expressions of the matter source densities~\eqref{sourcedensity} as a function of the PN potentials at the required order:
\begin{align}
& \sigma = \frac{2(d-2)}{\phi_{0}^{\frac{d-1}{2}}(d-1)}\,m_{1}\,\left[1+\frac{1}{c^{2}}\left((1-2s_{1})\left(\psi_{(0)}\right)_{1}+\frac{d}{2(d-2)}v_{1}^{2}-\frac{4-d}{d-2}(V)_{1}\right)\right]\delta^{(d)}\left(\mathbf{x}-\mathbf{y}_{1}\right) +[1\leftrightarrow 2]\,,\\
& \sigma_{i} = \frac{1}{\phi_{0}^{\frac{d-1}{2}}}\,m_{1}\,\left[1+\frac{1}{c^{2}}\left((1-2s_{1})\left(\psi_{(0)}\right)_{1}+\frac{1}{2}v_{1}^{2}-\frac{4-d}{d-2}(V)_{1}\right)\right]v_{1}^{i}\,\delta^{(d)}\left(\mathbf{x}-\mathbf{y}_{1}\right) +[1\leftrightarrow 2]\,,\\
& \sigma_{ij} = \frac{1}{\phi_{0}^{\frac{d-1}{2}}}\,m_{1}\,
v_{1}^{i}v_{1}^{j}
\,\delta^{(d)}\left(\mathbf{x}-\mathbf{y}_{1}\right) +[1\leftrightarrow 2]\,,\\
\nn & \sigma_{s} = \frac{2}{\phi_{0}^{\frac{d-1}{2}}(d(d-1)+4\omega_{0})}\,m_{1}\,\biggl[(1-2s_{1})+\frac{1}{c^{2}}\biggl(\Bigl((1-2s_{1})^{2}+4s'_{1}+\frac{4\phi_{0}\omega'_{0}}{d(d-1)+4\omega_{0}}(1-2s_{1})\Bigr)\left(\psi_{(0)}\right)_{1} \\
&\qquad\qquad -\frac{1}{2}(1-2s_{1})v_{1}^{2}-(1-2s_{1})(V)_{1}\biggr)\biggr]\delta^{(d)}\left(\mathbf{x}-\mathbf{y}_{1}\right) +[1\leftrightarrow 2]\,.
\end{align}




\bibliography{ListeRef_ST}

\end{document}